%% file: ms.tex
\documentclass[twocolumn]{emulateapj}
\usepackage[parfill]{parskip}         
\usepackage{graphicx}
\usepackage{amssymb}
\usepackage{epstopdf}
\usepackage{amsmath}   
\usepackage{amsthm}    
\usepackage{amssymb}   
\usepackage{graphicx}  
\usepackage{bm}
\begin{document}

\addtolength\topmargin{1cm}
\input{macros.tex}

\slugcomment{Accepted for publication in \apj}
\shortauthors{Nierenberg et al.}
\shorttitle{Cosmic evolution of substructure}

\title{Luminous satellites II:  Spatial Distribution, Luminosity Function and Cosmic Evolution}

\author{A.M. Nierenberg\altaffilmark{1}$^{*}$, M.W. Auger\altaffilmark{2}, T. Treu\altaffilmark{1}, P.J. Marshall\altaffilmark{3}, C.D. Fassnacht\altaffilmark{4}, 
Michael T. Busha\altaffilmark{5,}\altaffilmark{6}}
\altaffiltext{1}{Department of Physics, University of California, Santa Barbara, CA 93106, USA}
\altaffiltext{2}{Institute of Astronomy, University of Cambridge, Madingley Road, Cambridge, CB30HA}
\altaffiltext{3}{Department of Physics, University of Oxford, Keble Road, Oxford, OX1 3RH, UK}
\altaffiltext{4}{Department of Physics, University of California, Davis, CA 95616, USA}
\altaffiltext{5}{Institute for Theoretical Physics, University of Z\"{u}rich, Z\"{u}rich, Switzerland}
\altaffiltext{6}{Physics Division, Lawrence Berkeley National Laboratory, Berkeley, CA}
\altaffiltext{*}{{\tt amn01@physics.ucsb.edu}}


\begin{abstract}
We infer the normalization and the radial and angular distributions of
the number density of satellites of massive galaxies
($\log_{10}[M_{h}^*/M\odot]>10.5$) between redshifts 0.1 and 0.8 as a
function of host stellar mass, redshift, morphology and satellite
luminosity.  Exploiting the depth and resolution of the COSMOS HST
images, we detect satellites up to eight magnitudes fainter than the
host galaxies and as close as 0.3 (1.4) arcseconds (kpc). Describing
the number density profile of satellite galaxies to be a projected
power law such that $P(R)\propto R^{\rpower}$, we find
$\rpower=-1.1\pm 0.3$. We find no dependency of $\rpower$ on host
stellar mass, redshift, morphology or satellite luminosity. Satellites
of early-type hosts have angular distributions that are more flattened
than the host light profile and are aligned with its major axis. No
significant average alignment is detected for satellites of late-type
hosts.  The number of satellites within a fixed magnitude contrast
from a host galaxy is dependent on its stellar mass, with more massive
galaxies hosting significantly more satellites. Furthermore, high-mass
late-type hosts have significantly fewer satellites than early-type
galaxies of the same stellar mass, possibly indicating that they
reside in more massive halos. No significant evolution in the number
of satellites per host is detected. The cumulative luminosity function
of satellites is qualitatively in good agreement with that predicted
using subhalo abundance matching techniques. However, there are
significant residual discrepancies in the absolute normalization,
suggesting that properties other than the host galaxy luminosity or
stellar mass determine the number of satellites.
\end{abstract}

 ---------------------------------------------------------------------------
                      
\section{Introduction}
\label{sec:intro} 
Cold Dark Matter  (CDM) simulations of structure formation successfully match observations of both the  spatial and mass distribution of super-galaxy scale structure. 
However, they are less successful at matching observations of structure at smaller scales. 
For example, these simulations over-predict the number of low mass satellite companions of Milky Way mass halos \citep{Kauffmann++93, Klypin++99,Moore++1999,Strigari++07}. 
This discrepancy may demonstrate a fundamental breakdown of the $\Lambda$CDM paradigm at small scales, possibly indicating that assumptions about the nature of the CDM particle are incorrect. For instance, DM particles may have higher kinetic energy than expected \citep[e.g.][]{Colin++00, Schneider++11,Menci++12} leading to suppressed fluctuations in the primordial power spectrum at small scales, thereby decreasing the number of present day satellites \citep[e.g.][]{Kam++00, Zen++03}. 
Alternatively, the missing satellite problem may be due to the difficulty of observing very faint satellites due to low star formation efficiency in low mass halos. 
A variety of baryonic processes can produce suppressed star formation in low mass halos \citep[e.g][]{Thoul++1996,Gnedin++00,Kaufmann++08, Maccio++11, Springel++10}. However,
it is not easy to observationally confirm which processes have the strongest effects, in part due to the small number of faint satellite galaxies studied outside of the Local Group. 

In the Local Group, the luminosity function of satellite galaxies has been studied to a faintness of $M_V = -2$ \citep[see][and references therein]{Koposov++08}. Outside of the Local Group, measurements have been restricted to brighter satellites. Multiple studies using data from SDSS have measured the luminosity function of satellites at low redshift \citep{Guo++11,Liu++11,Lares++11,Strigari++11}. The deepest study by \citet{Guo++11}, measured the luminosity function reliably for satellites up to 7.5 magnitudes fainter than the hosts, while \citet{Strigari++11} placed upper limits on satellite numbers for satellites up to 10 magnitudes fainter. All of these studies agreed on two main conclusions. 
The first was that the luminosity function of satellites is dependent on the host luminosity. This is due to the fact that for massive host galaxies, the host stellar mass is believed to be a non-linear function of the host virial mass \citep[e.g.][]{Behroozi++10,Wake++11,Leauthaud++11}. Thus while the number of satellites at a fixed \emph{virial} mass ratio with the host is expected to be scale invariant \citep{Krav++10}, the luminosity ratios may not be. 
The second conclusion from these studies was that while the number of faint satellites appears to be consistent with that in the Local Group down to an absolute V band magnitude of -14 \citep{Guo++11, Lares++11,Strigari++11}, on the bright end only 10-20 percent of hosts with Milky Way and Andromeda luminosities host satellites as luminous as the Large and Small Magellanic Clouds (L/SMCs). The rarity of high mass companions is consistent with predictions from  $\Lambda$CDM simulations, indicating that the Milky Way luminosity function is an outlier on the bright end rather than that there is an issue with the theory.  
 However, a significant discrepancy between theoretical predictions and observation of the satellite luminosity function only begins to appear for satellites with absolute magnitudes fainter than $M_{V} > -14$. Thus, in order to test whether the missing satellite problem is due to the Milky Way having an anomalous lack of satellites on the faint end, it is necessary to study even fainter satellites.

In addition to providing a key to understanding star formation in low mass halos, and potentially allowing us to constrain the mass of the DM particle,
 satellite galaxies are believed to have played a significant role in the evolution of the size of the most massive galaxies ($\Mstar >10^{10} \Msun$)
 \citep{Boylan-Kolchin++06, Boylan-Kolchin++07, Hopkins++10}\footnote{But see also \citet{Nipoti++12}}. The role of minor mergers in the evolution in size and mass has been observed in numerous observational 
 studies which used pair counting and assumptions about merger time-scales to estimate the minor merger rate \citep{LeFevre++00, Bell++06, Patton++08, Bundy++09, Bezanson++09,Robaina++10,Newman++11,Tal++11}. 
 Furthermore,  measurements of disturbed morphology and color gradients in an elliptical hosts as evidence for recent mergers \citep{Kaviraj++09, Kaviraj++11}
  show that minor merging is a key contributor to low level star formation seen in the outskirts of early-type galaxies.
 These studies have all been limited to the study of the most massive companions with stellar masses at least ten percent of their host mass, 
 which is ten times higher than the mass ratio between the Milky Way and the LMC. 
In order to better constrain the effect of minor mergers on the evolution of 
massive galaxies, it is necessary to push the study of companions at cosmological distances to lower stellar masses  
\citep[see also][]{Bundy++07, Naab++09, Fakhouri++10}. 
 
The spatial distribution of satellites about their host galaxies is intimately tied to the luminosity function of those satellite galaxies. 
Two competing effects contribute to the link between
satellite luminosity function and spatial distribution. On the one hand, baryons are thought 
to play an important role in the preservation of subhalos in the inner regions of the host dark matter halo by steepening the total mass profiles  and making them less susceptible to tidal stripping. Numerous hydrodynamical and semi-analytic simulations have found that subhaloes with baryons are more centrally located in the host dark matter halo than pure dark matter subhalos \citep[e.g.][]{Blu++86,Libeskind++10,Romano++09,Weinberg++08,Maccio++06,Gao++04}.

At the same time, baryons will also steepen the mass profile of the host, increasing the strength of tidal shock heating of the gas and stars in satellites as they pass near the center of the 
halo, with the amount of heating and mass loss strongly dependent on the orbits of the satellites \citep{Gnedin++99,Choi++09, Boylan-Kolchin++07,Dolag++09}. \citet{Donghia++10} find that the effects of these interactions in suppressing star formation is stronger for less massive subhalos.
For these reasons, the slope of the radial profile of satellites is of key importance to any analysis attempting to recover the physics that governs the interactions between satellites and their host halo and central galaxy.

The radial profile of the number density of satellites has been measured with somewhat contradictory results. 
Assuming a single power-law model for the projected number density of satellites $N$ as a function of distance from the host, where $N(R)\propto R^\gamma_p$, \citet{Chen++08} measured $\gamma_p = -1.5 \pm 0.07$ for satellites brighter than $M_r \sim -17.5$ in SDSS. 
In a study of SDSS satellites brighter than $M_g <-21.2$, \citet{Watson++10} measured $\gamma_p = -1.2 \pm 0.1$. In \citet{Nierenberg++11} (hereafter \paperone), we measured the slope of higher redshift satellites up to 5.5 magnitudes fainter than their hosts (on average about $M_{V}<-16$) to be $\gamma_p = -1.0^{+0.3}_{-0.4}$.

 There are several likely explanations for the discrepancies in these measurements. One possibility is differences in the technique 
used to separate satellites from background/foreground interlopers. Alternatively the difference in inferred slopes may be due to differences in host masses studied, or differences in the luminosity of the satellites. 
For example, \citet{Watson++11} measured the radial profile of satellite galaxies brighter than $M_r < -18$ between 7 and 280 kpc as a function of satellite luminosity, and found that the faintest satellites had significantly shallower radial profiles than their more luminous counterparts. \citet{Tal++12} also found fewer faint satellites in groups within 25 kpc for satellites brighter than $M_g<-22.8$. However outside of this region, they found that all satellites followed the same radial profile which was well described by a combined Sersic+NFW profile, mimicking the total mass profile which follows $R^{-1}$ \citep{Gav++07, Aug++10} for massive galaxies. In contrast to these studies, \citet{Guo++12} find that for satellites brighter than $m_r<22$ in SDSS, fainter satellites are \emph{more} centrally concentrated than bright satellites. \citet{Budzynski++12} also found an excess of fainter satellites in the innermost regions for satellites in groups and clusters.

To make progress on understanding the satellite radial profile, it is necessary to make measurements that take into account systematics such as the host mass and satellite luminosity. In addition, in order to learn about trends in merging rates it is interesting to study evolution in the radial profile. \citet{Budzynski++12} found no significant evolution in the radial profile of satellites in groups and clusters between redshifts 0.15 and 0.4. In this work we extend the measurement of the satellite radial profile to fainter satellites and to higher redshifts.

The angular distribution of satellites also contains information about
the effects of anisotropic, filamentary accretion in $\Lambda$CDM and
the shape of the host dark matter halo. Pure dark matter simulations
predict that satellites should be found in an anisotropic distribution
which is aligned with the major axis of the host dark matter halo
\citep[e.g.][]{Zentner++05, Zentner++06,
Knebe++04,Aubert++04,Faltenbacher++07,Faltenbacher++08}.  Thus by
studying the orientation of satellite galaxies with respect to the
major axis of the host \emph{light profile} it may be possible to
learn about the relative orientation between the stars in the host
galaxy and the dark matter halo they reside in.  In the Milky Way and
Andromeda, satellite galaxies appear in an anisotropic distribution,
preferentially aligned with the minor axis of their hosts
~\citep[e.g.][]{Metz++09}. When averaged, however, studies of
satellites in SDSS find that satellites are randomly located around
late-type galaxies \citep{Bailin++08,Yegorova++11} and aligned with
the major axis of the host light profile for early-type hosts
\citep{Brainerd++05,Agustsson++10}.  The studies of the satellite
angular distribution outside of the Local Group have all been limited
to relatively bright satellites (within 1-2 magnitudes of the host
magnitude), and they have not attempted to measure the relative
ellipticity of the satellite angular distribution.  As stated above,
measuring the angular distribution of fainter satellites is of key
interest because several simulations predict that satellites that have
very elongated orbits will lose the most gas as they pass through the
host galaxy.

Finally, the measurement of the luminosity function and radial profile of satellites at intermediate redshifts can be combined with gravitational lensing studies to constrain star formation efficiency in subhalos \citep[][and references
therein]{Treu++10, Krav++10}. Recently low mass subhalos have been detected in mass reconstructions of gravitational lenses \citep{Vegetti++12, Vegetti++10b,McK++05,MacLeod++09}. By analyzing the probability of detecting satellites of these masses near the lensed images, \citet{Vegetti++12} constrained the low mass end of the mass function of subhalos to have a slope of $\alpha = 1.1^{+0.6}_{-0.4}$. 
At present, the combination of gravitational lensing analyses with luminosity function measurements is the only method to directly measure stellar to virial mass ratios for satellites outside of the Local Group.

In this paper we address three questions: 1) What is the spatial distribution and cumulative luminosity function of the satellites of massive galaxies? 2) How does the spatial distribution vary with satellite luminosity? 3) How do these properties vary with host morphology, stellar mass and redshift? 
We conclude by discussing how our results can be interpreted in the context of $\Lambda$CDM.

In \paperone, we developed a method of modeling and subtracting the host light profile to allow us to detect faint ($m_{sat}-m_{host}>5.5$) satellites at intermediate redshifts near early-type host galaxies in the GOODS  \citep{Gia++04} fields. We also introduced a statistical model which we used to simultaneously infer the spatial and angular distribution of satellites along with their numbers. In this work, we expand our analysis to include the satellites of host galaxies selected from the COSMOS \citep{Scov++07} field which is approximately 20 times larger than the GOODS field although somewhat shallower. 
This dramatic increase
in our sample size allows us to analyze the properties of the 
radial and angular distribution and the number of satellites, 
for satellites more than a thousand times fainter than their host galaxies. 
Furthermore, the greatly increased number of host-satellite systems allows us to 
analyze these properties in bins of redshift, host morphology, satellite luminosity and host mass. 
This is of key importance because, as has been discussed, the number and luminosity of satellites are believed
to depend strongly on the properties of host galaxies.
In order to make our results more easily relatable to theoretical predictions, we select our host samples by stellar mass and luminosity and analyze the number density of the satellite radial profile in units of $\rtwo$ in addition to units of the half-width at half-max of the host light profile.

This paper is organized as follows: In \S \ref{sec:data}, we describe
the images and catalogs used in this analysis. 
In \S \ref{sec:sample} we discuss
the selection and properties of the host galaxy sample. In \S \ref{sec:subtraction}
we briefly review the host modeling and subtraction
technique developed in \paperone. 
 In \S \ref{sec:firstlook} we present the binned radial and angular profiles
 of satellites to provide a qualitative sense of the signal. 
 In \S \ref{sec:model} we review the methodology for
a full Bayesian statistical analysis of the data developed in \paperone~which allows us to 
use a catalog of object positions and magnitudes to infer the presence of satellites
against the homogeneous signal of background/foreground objects.
In \S \ref{sec:theornum} we discuss the theoretical model of the cumulative number
of satellites per host as a function of host luminosity from \citet{Busha++11} which 
we compare with our measurements.
In \S \ref{sec:analysis} we explain how we tested our statistical analysis and its limitations.
In \S~\ref{sec:results} we present the results. In \S~\ref{sec:comparison} we compare
our results with similar studies performed at low redshift. 
In \S~\ref{sec:discussion} we discuss the broader implications
of our results. In \S~\ref{sec:summary} we provide a concise summary. The
Appendix contains more detailed explanations of many of the methods 
used in this paper.

Throughout this paper, we assume a flat $\Lambda$CDM cosmology with
$h=0.7$ and $\Omega_{\rm m}=0.3$.  All magnitudes are given in the AB
system \citep{Oke++1974} unless otherwise stated.

\section{Imaging and Catalogs}
\label{sec:data}
To achieve the depth and area required to study satellite properties and their evolution, we use imaging and photometric catalogs taken from the COSMOS survey \footnote{COSMOS photometric catalogs and surveys are available at \text{http://irsa.ipac.caltech.edu/data/COSMOS/datasets.html}}.
 In \paperone~we used the deeper GOODS survey to develop and test our method. However, the GOODS survey is not sufficiently wide to allow us to perform a binned statistical analysis of the satellite population as a function of host properties as we do in this paper. 
 
The COSMOS field has extensive spectroscopic and photometric ground-based coverage which we use to catalog the 
redshift, stellar mass and morphology of the host galaxies.
Spectroscopic redshift measurements from \citet{Lilly++07} are available for roughly 
1/7 of the host galaxies. 
When spectroscopic measurements are not available we use ground-based photometric redshift measurements from the catalog by \citet{Ilbert++09}, 
which is the same catalog we use for all stellar mass measurements. 
Finally for morphological categorization, we use the \citet{Cassata++07} morphological catalog.
The \citet{Cassata++07} catalog is created by a computer algorithm 
which is known to fail in certain cases.
For instance, spiral galaxies with large bulges can be misclassified as ellipticals. Thus we
visually confirmed all morphological classifications using COSMOS 
ACS imaging. This resulted in the re-categorization of $\sim 10\%$ of the morphologies of
host galaxies.

\section{Host Galaxy Selection}
\label{sec:sample}
In selecting our host galaxy sample, we had two main goals. The first was to ensure that we selected hosts that maximized our ability to detect neighboring satellites, and the second was to select a host sample that could be easily related to predictions from simulations. In \paperone, we achieved the first goal
by selecting massive early-type host galaxies which had relatively smooth light profiles that made it easy to find nearby companions. However, galaxy morphology is not well reproduced by current simulations \citep[e.g.][]{Hoyle++11}, making it difficult to select matching host galaxies from simulations.
In contrast, simulations have had success matching the stellar mass function of massive galaxies, to 
dark matter halos in simulations \citep{Berlind++02,Zheng++07,Conroy++09}, making a mass-selected host sample more appealing.

To compromise between the two goals, we selected relatively high stellar mass $\log[M^{*}_{\rm h}/\Msun]>10.5$ host galaxies, as these have well studied stellar masses, redshifts and luminosities, and a greater number of satellites visible down to a fixed apparent magnitude \citep[e.g.][]{Krav++10,Busha++11}. 
We study hosts in a redshift range between $0.1<z<0.8$. This range
allows us to study evolutionary properties while still
guaranteeing that we can detect
satellites with luminosity contrast from the hosts
equivalent to that of the SMC relative to the Milky Way at all redshifts. 
By doing this, we expect to observe approximately one satellite per host. 
We exclude $z<0.1$ galaxies, which are few and too extended in angular size to
analyze in the same way as the more distant sample.

We also required host galaxies to be relatively isolated in order to ensure that they themselves are not satellites of larger central galaxies. 
This is important for our analysis because we do not want to count objects associated with the larger host as satellites of a smaller galaxy. 
Using the stellar mass and redshift catalogs, we include only host galaxies which are not within the $\rtwo$ of a neighbor that has more than its stellar mass and is at the same redshift within measurement uncertainties.We calculate $\rtwo$ given stellar mass using Equation 3 from \citet{Dutton++10}, which provides a by eye fit to the observed and inferred relationships between stellar and halo mass of galaxies as a function of morphology. For early-type galaxies the best fit function is:
\be
y = 10^{2.0}\left(\frac{x}{10^{10.8}}\right)^{-0.15}\left[\frac{1}{2} + \frac{1}{2}\left(\frac{x}{10^{10.8}}\right)^{2}\right]^{0.5}
\ee
And for late-types:
\be
y = 10^{1.6}\left(\frac{x}{10^{10.4}}\right)^{-0.5}\left[\frac{1}{2} + \frac{1}{2}\left(\frac{x}{10^{10.4}}\right)\right]^{0.5}
\ee
Where $y = \left<M_{200}\right>/M^*$ and $x = M^*$.

\begin{figure*}[h!]
\centering
\includegraphics[scale=.4, trim = 0 0 50 200,clip =true]{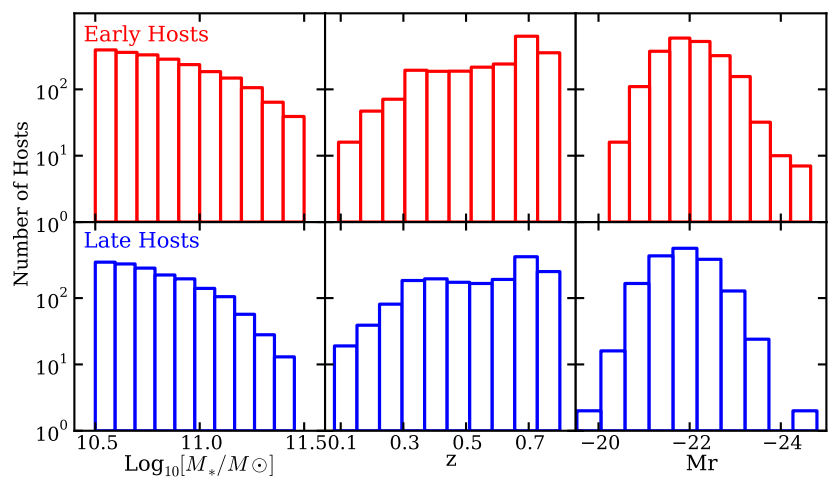}
\caption{The distribution of stellar mass, redshift and absolute r band magnitude for \emph{upper}: early and \emph{lower}: late-type hosts.}
\label{fig:host_redshifts}
\end{figure*}

Although there is uncertainty in the stellar mass estimate and additionally in $\rtwo$, we consider these values to be know with absolute precision for the purposes of our analysis. We tested the effect of uncertain stellar mass on our analysis by doubling masses before calculating $\rtwo$ and found no significant impact on our inference result. The final sample has 1901 early-type and 1524 late-type galaxies. The distribution of stellar masses, redshift and absolute r-band magnitudes are shown in Figure \ref{fig:host_redshifts}. 

As discussed in the Introduction, there is theoretical and observational evidence  that satellite properties may depend strongly on the properties of host galaxies.  In order to study these trends,  we divide host galaxies into bins of redshift, morphology and stellar mass when performing our analysis of the satellite population. We choose two bins of stellar mass with $10.5<\log_{10}M^{*}_{\rm h}<11.0$ and $11.0<\log_{10}M^{*}_{\rm h}<11.5$, and two bins in redshift with $0.1<$z$<0.4$ and $0.4<$z$<0.8$. 

When we analyze the satellite population, we study bins of cumulative magnitude contrast from the host galaxy, such that all satellites have magnitudes brighter than $m_{\rm sat}-m_{\rm host}<\dm$. In each bin of $\dm$ we only study host-satellite systems such that the host is at least $\dm$ magnitudes brighter than the limiting survey magnitude. This means a different subset of host galaxies is used to study $\dm = 8.0$ satellites than $\dm = 2.0$ satellites, leading to some variation in host properties within a stellar mass, redshift and morphology bin. In Tables \ref{tab:modResultsEarly} and \ref{tab:modResultsLate}, we summarize the average stellar mass, redshift and luminosity of the host galaxies used in each $\dm$ analysis, split into bins of redshift, stellar mass and host morphology. Host galaxies used to study the faintest satellites tend to be at lower redshift than the average in a fixed redshift bin.

\section{Detection and photometry of close neighbors}
\label{sec:subtraction}
 
In order to study the satellite population, we require an accurate catalog of object positions
and magnitudes as close as possible to the host galaxies.
Companions of bright galaxies are difficult to study because
they are intrinsically faint and often obscured by the host galaxy
light.  This is a serious issue when attempting to measure the slope of 
the power law-radial profile of the satellite spatial distribution, as the innermost
regions of the system provide the best constraint on the slope. Furthermore, close to the host, the ratio of the number
of satellites to background/foreground galaxies is the most favorable. 
In \paperone, we developed a method of removing the smooth component of the host galaxy light
profile to overcome some of these challenges. 
This process makes automated object detection much more accurate 
and reliable near the host galaxy. We use the results to update the COSMOS photometric catalog
with newly detected objects, and to replace the photometry for objects that had already been detected
 near the hosts. 
 
In this section we briefly review the method of host light
modelling and subtraction developed in \paperone~and we discuss the 
new objects this method allows us to identify. All \sextractor parameters used are listed in 
Table \ref{table:cos_SePars} in the Appendix.

\subsection{Companion masking and host light subtraction}  
Host subtraction allows us to accurately identify faint objects near our host galaxies using similar \sextractor \citep{Bertin++1996} parameters to those that were to make the COSMOS photometric catalogs. 
These parameters were conservative as they were intended to study objects across a large range of angular sizes and morphologies. To ensure that we are able to identify faint objects that the conservative parameters might have otherwise missed, we remove the light from the images in a three step process.

 In the first step, we use  \sextractor \citep{Bertin++1996} to identify and then mask objects in a small cutout region of 10 $\Rhost$ near the host galaxy, where $\Rhost$ is the second moment of the host galaxy light profile along its major axis (A\_IMAGE). 
  In this step, we select more `liberal' \sextractor parameters than are used in the final round of object detection, erring on the side of masking noise, to ensure faint satellites are masked. Two separate rounds of object identification are combined to make the final mask, one which is tuned to detect diffuse objects farther from the central galaxy and the second which is tuned to detect faint compact objects closer to the central galaxy.  Finally, we use a two dimensional, elliptical b-spline model to model the masked image and subtract the host light profile. The model we use is similar to that described by \citet{Bol++05,Bol++06}.

\subsection{Object detection and photometry}

We detect objects in the host-subtracted images using \sextractor parameters tuned to match the object detection and photometry that was used to make the COSMOS catalogs. Appendix \ref{app:SE} contains a comparison of our photometry and object detection to that of the COSMOS catalogs in a large non host-subtracted field. 

In order to test our sensitivity to low surface brightness objects, we simulated faint sources near our hosts with Sersic indices and effective radii and tested our recovery rate and photometric accuracy for these objects after host subtraction. Results of these simulations can be found in Figure \ref{fig:completeness} in the Appendix. 
We use the results from these simulations to identify a minimum radius at which we can detect at least 90 percent of simulated satellites with accurate photometry, and define this as the minimum radius at which we study the properties of the satellite population. We find that we can accurately recover satellites with  MAG\_AUTO $\mi< 25.0$ as close as 2.5 \Rh~ in COSMOS for the majority of hosts with early-type light profiles. This corresponds to a mean distance of 1\farcs2 (7 kpc) with a standard deviation of 0\farcs7 (3 kpc). 
The inner detection boundary is slightly higher for host galaxies with $\MstarHost>11.0$ and $z<0.4$ as these tend to have light profiles that extend above the background further from the host centers.
Late-type galaxies have more extended light distributions in addition to spiral arms which are difficult to distinguish from neighboring galaxies, thus we choose a more conservative inner boundary of 4 \Rh  (3$\pm$1 arcseconds, 17$\pm$ 6 kpc) for these hosts.  On average, these minimum radii correspond to 0.02 and 0.07 $\rtwo$ for early and late-type galaxies respectively.

\subsection{Properties of Objects Detected in Cutout Regions}
\label{subsec:newObjects}
\begin{figure}[h!]
\centering
\includegraphics[scale = 0.35]{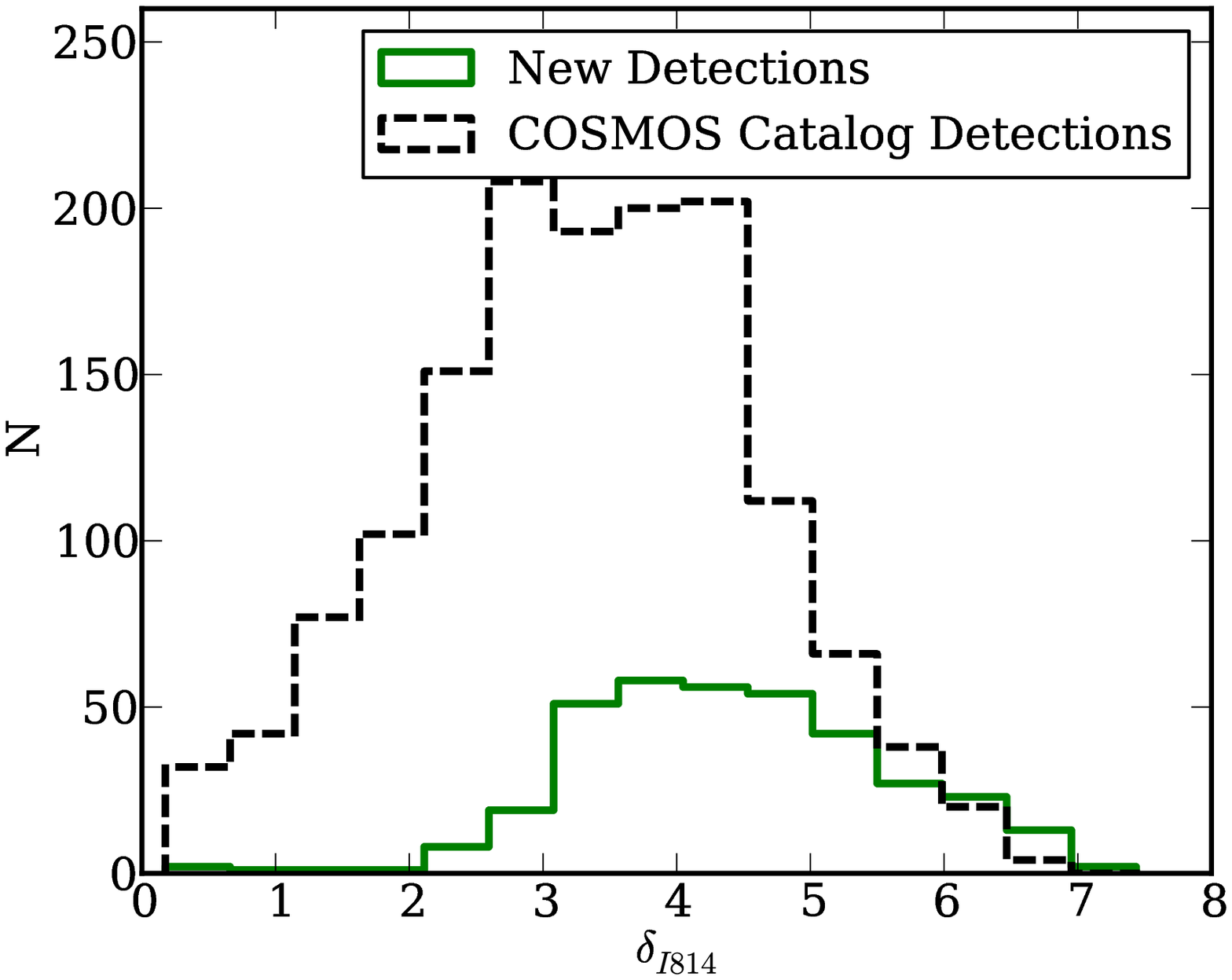}\\
\includegraphics[scale = 0.35]{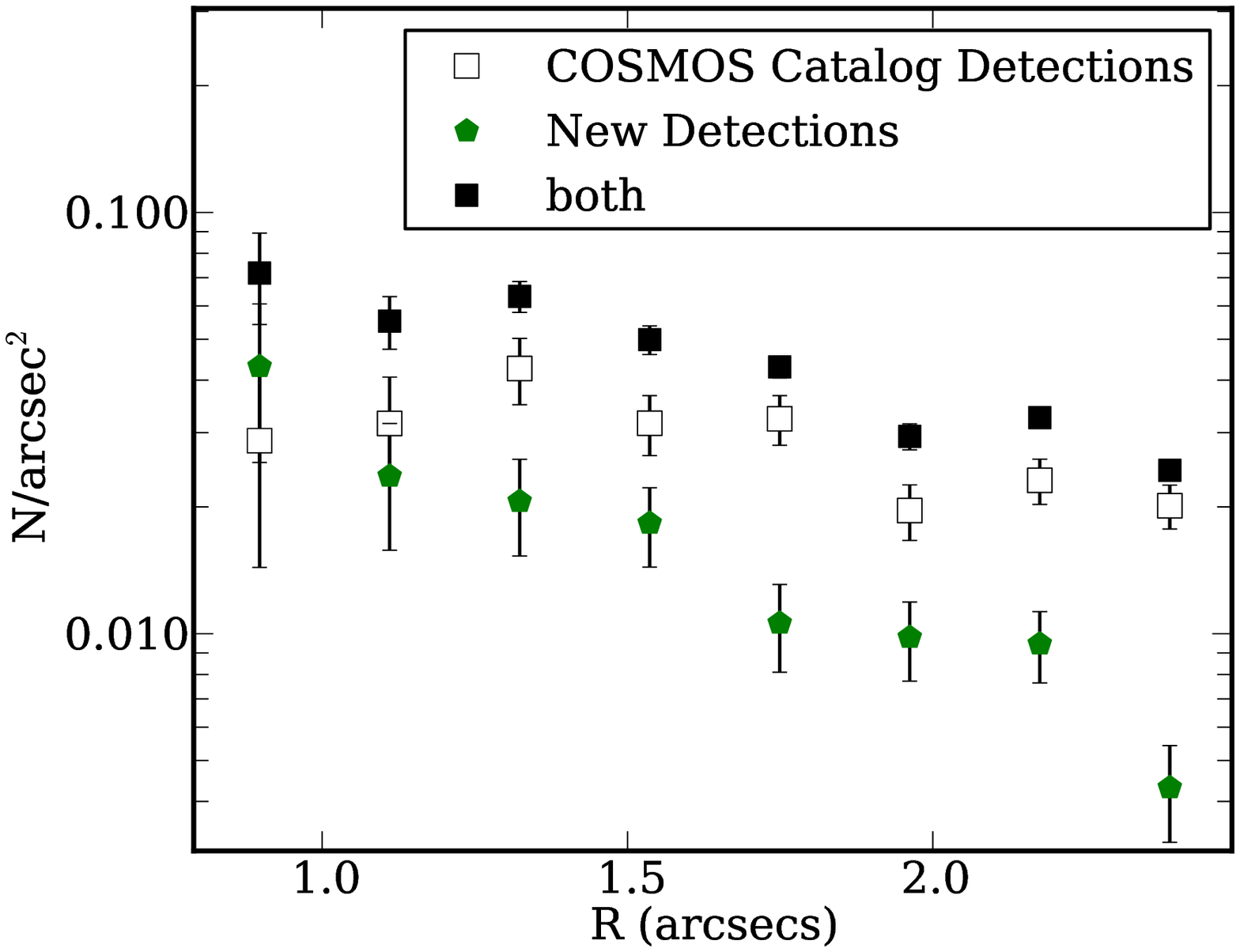}
\caption{Comparison of the properties of objects detected in the COSMOS catalogs to those of newly detected objects after host light subtraction. \emph{Upper}: The distribution of magnitude differences from hosts ($\delta$m$ = m - m_h$) within 2\farcs5. \emph{Lower}: The number density of objects as a function of distance from hosts. Newly detected objects are closer to the hosts than those in the COSMOS photometric catalog and make a significant contribution to the measurement of object number density within 2 arcseconds from the center of the host.}
\label{fig:mus}
\end{figure}

In this section we compare the properties of newly detected objects after host light subtraction
to the properties of objects already in the COSMOS photometric catalog near the host galaxies.
The upper panel of Figure \ref{fig:mus} shows the distribution in the 
contrast in MAG\_AUTO measurement ($\delta$m$ = m - m_{\rm h}$)\footnote{Note the use of lower-case $\delta$m, which denotes a
specific contrast from the host and is different from $\dm$ which
describes the allowed maximum contrast between host and neighboring
objects for a particular data set.} between hosts
and objects detected within 2\farcs5 of the host galaxies. The distributions are compared for
objects already in the COSMOS catalog and newly detected objects. For both types of objects,
 photometry is performed after host light removal.
Newly detected objects are about a magnitude 
fainter than objects that were already in the COSMOS photometric catalog, with average 
$\delta$m values of 3.22 compared to 4.44 for previously detected objects within the same region, 
with typical measurement uncertainty of 0.05 mag. 

The lower panel of Figure \ref{fig:mus} shows the number density of objects as a function of distance from the hosts. The number density 
of objects in the COSMOS photometric catalogs drops within the inner 1\farcs5, while the number density of newly detected objects rises, more than doubling the number
of COSMOS detected objects in this region. The sum of the two number density signals increases steadily with decreasing distance from the host galaxy. Thus host subtraction and rigorous measurements of completeness in the innermost regions are necessary for an accurate measurement of the radial profile and number of objects close in projection to the host galaxies. 

\section{First Look}
\label{sec:firstlook}
Before describing our model for the radial and angular profiles of
objects near the host galaxies, it is instructive to show these distributions
in spatial bins in order to provide a visual representation of the
data. However, binning is inherently limited because it requires the averaging
of data, thereby losing information. Furthermore it is not conducive to accurate
subtraction of foreground/background galaxies over a range of redshifts.  Thus we do
not perform our analysis on the spatially binned data, but instead use this
section to justify our model choices in Section \ref{sec:satmod}.

\subsection{Distance Scaling and Radial Distribution}
\label{sec:BinnedRadii} 

We scale measured object distances to account for the range of redshift and host mass scales in our sample. 
A scale relating to the host light profile is the natural choice for the observer as this will vary
with host redshift as well as host mass according to the size-mass relation \citep[e.g.][]{Tru++06,Williams++10}. 
For this distance scale, we use $\Rhost$ which is AWIN\_IMAGE from \sextractor. 
We also perform a parallel study with all distances
scaled by $\rtwo$ of the host galaxies.  Unlike $\Rhost$, $\rtwo$ can be calculated in dark matter only simulations
and is thus a better choice when attempting to compare results with simulations.
However, estimating $\rtwo$ requires multiband photometry and stellar mass modeling which is not always possible, so it is useful 
to perform this analysis using both distance scalings to see if one choice or the other leads to systematic differences.
 
\begin{figure}[h!]
\centering
\includegraphics[width = 3.35in,height = 2.5in]{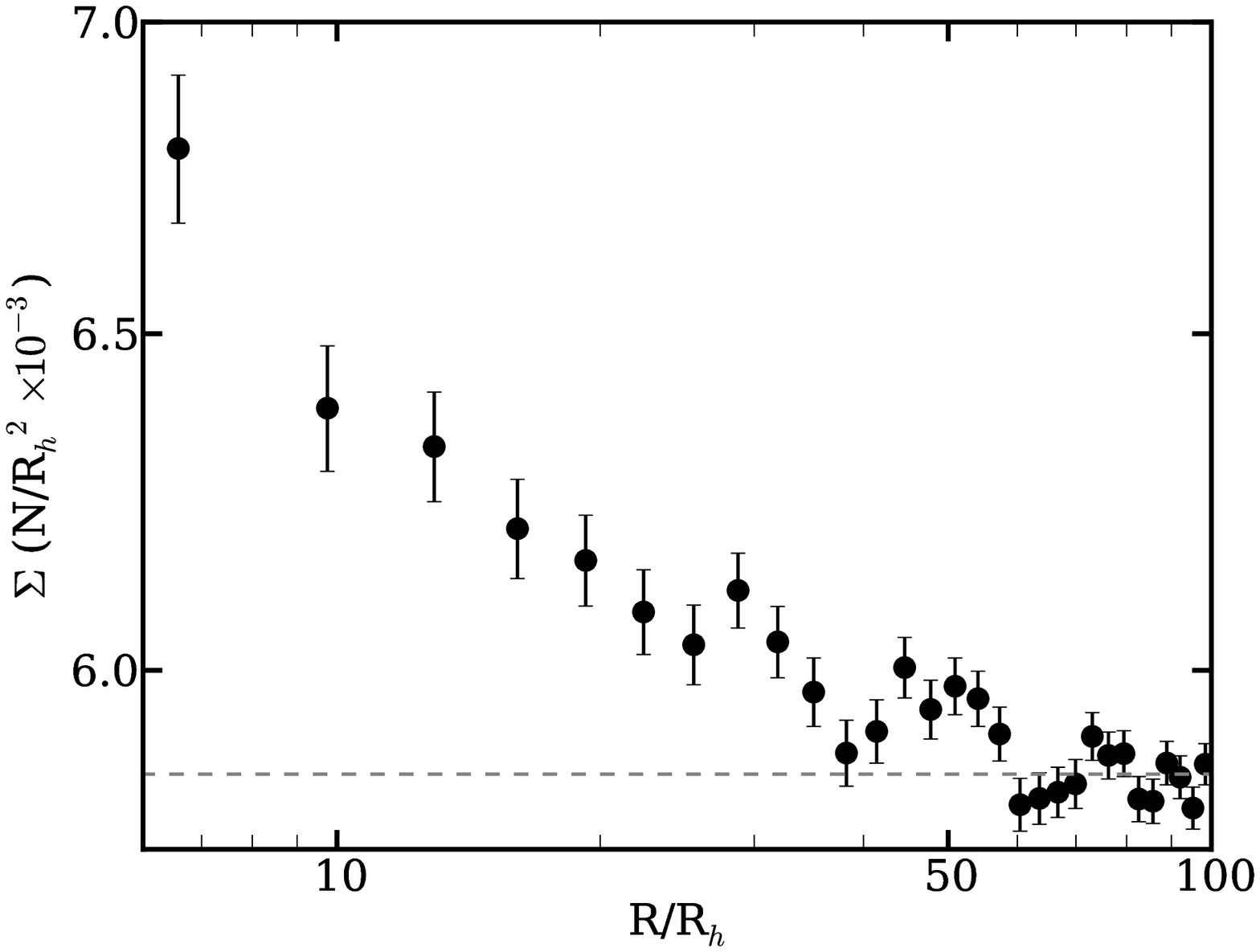} \\
\includegraphics[width = 3.35in,height = 2.5in]{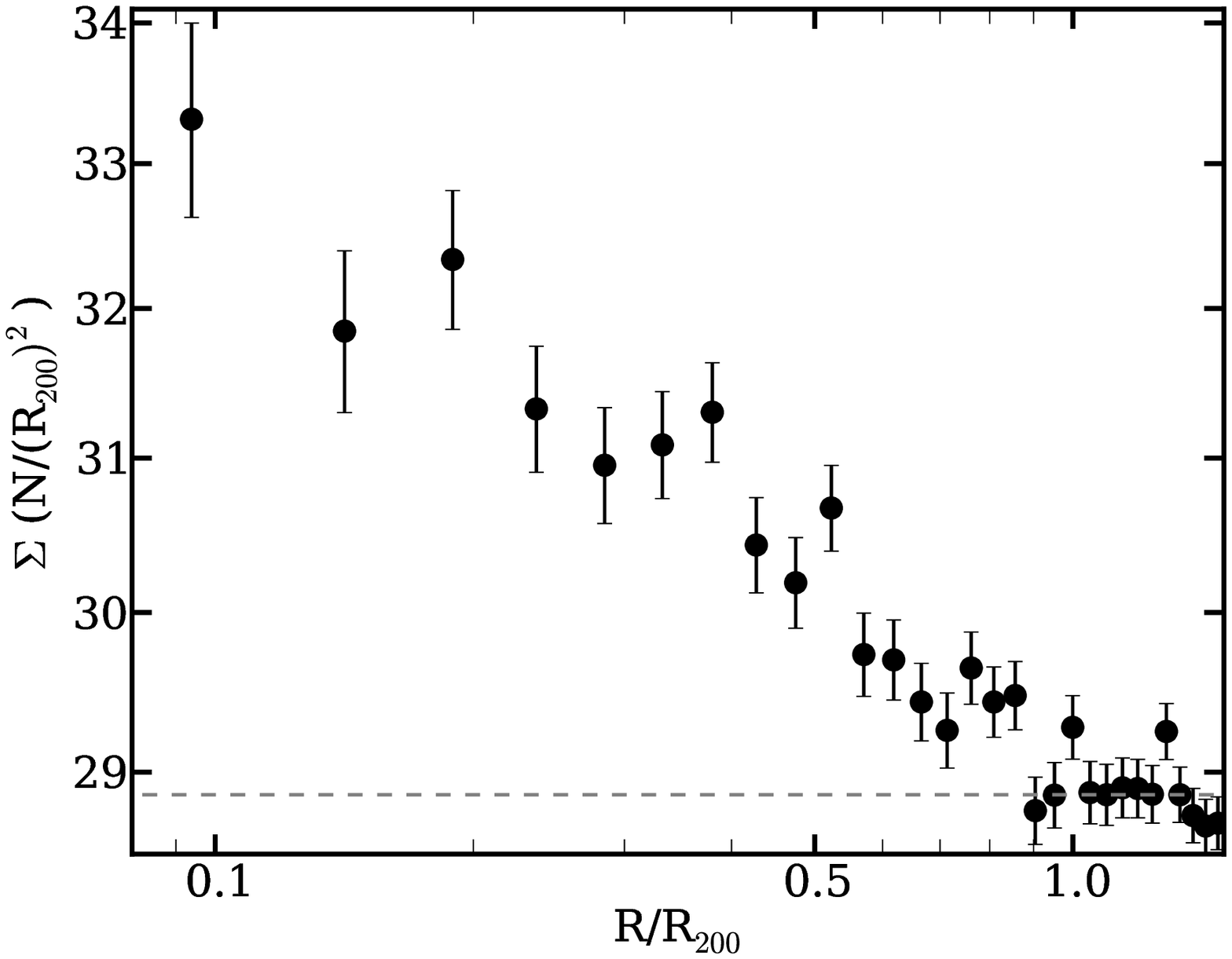}
\caption{The average number density of objects near hosts as a function of radial distance. \emph{Upper}: In units of the second order moment of the intensity of the host-light profile along its major axis ($\Rhost$), and \emph{lower}: In units of $\Rvir$ estimated from the host stellar mass.} 
\label{fig:prs}
\end{figure}

Figure \ref{fig:prs} shows the average number density of objects as a function of distance from
the hosts, with distances scaled by $\Rhost$ in the upper panel and $\rtwo$ in the lower panel.
The behavior is qualitatively similar for both choices of distance scaling;
the number density of sources increases as a power-law near the hosts. 
At large radii, the number density becomes dominated by the isotropic and
homogeneous distribution of objects not associated with the hosts, represented by the gray dashed lines. 

In  Section \ref{sec:model} we describe how we analyze the number density
signal by inferring the combined properties of the satellite and background/foreground 
populations. In Section \ref{sec:results_gammas}, there is a comparison of the results using the two
distance scalings. 

\subsection{Angular Distribution}
\begin{figure}[h!]
\centering
\includegraphics[scale=.6, trim = 0 0 300 0,clip =true]{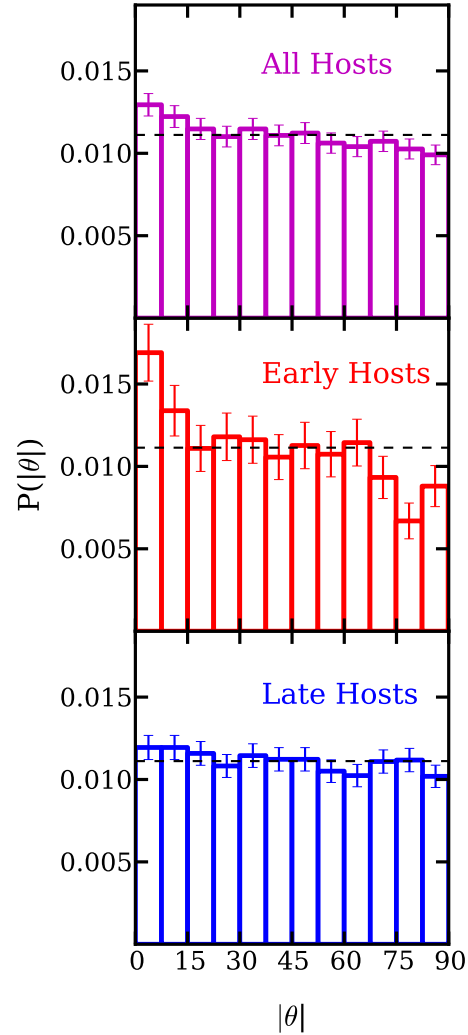}\\
\caption{The average number of objects at angle $|\theta|$ from the major axis of the host light profile, within 10 $\Rhost$, for hosts with elongation (b/a) less than 0.6. The distribution of objects near early-type hosts is more aligned with the host light profile than the distribution of objects near late-type hosts.} 
\label{fig:pthetas}
\end{figure}

In Figure \ref{fig:pthetas} we show the angular distribution of objects within 10 $\Rhost$, plotted for all hosts, early-type hosts and late-type hosts,
 where $\theta=0$ is aligned with to the major axes of the host light profiles. This figure only includes host galaxies with axis ratio b/a$<$0.6, to ensure that the direction of the host major axis is clearly measurable. 
 As background/foreground objects are expected to be distributed isotropically relative to the host galaxy, any anisotropy we observe is caused by correlated structure presumably, in the form of satellites. In this region we expect a significant contribution to the number density to come from satellites, as evidenced by the strong satellite signal within this region in the upper panel of Figure \ref{fig:prs}.

 Near early-type galaxies, the angular distribution shows a dominant component aligned with $\theta = 0$. A Komogorov-Smirnoff KS test gives a probability of $\sim 10^{-8}$ that the objects near early-type galaxies have a uniform angular distribution. In contrast, the objects near late-type galaxies appear more isotropically with a KS probability of being  uniform of a few percent. This simple examination has been done without any effort to remove background/foreground contamination. However, it indicates that the satellites of early and late-type hosts may have different angular distributions relative to their host light profiles. To test this further, we separate satellite populations based on host morphology in our inference of the parameters of the satellite spatial distribution in Section \ref{sec:satmod}.

\section{Joint Modeling of Satellite and Background Galaxy Populations}
\label{sec:model}
In this section we review how we infer the number of satellites and their radial and angular distributions from
a catalog of object positions and magnitudes. The basic procedure is to model the 
 observed positions and number density of objects  as 
a combined signal from background/foreground objects and satellite galaxies. These two populations
have significantly different spatial distributions; satellite galaxies increase in number near the host galaxies
and for early-type galaxies appear preferentially aligned with the major axis, whereas
background/foreground objects have a homogeneous and isotropic number density signal. 
By inferring the properties of the combined signal, and using prior information about the background/foreground
objects, we isolate the satellite number density signal. 

In Subsection \ref{sec:satmod} we discuss the details of this model for the satellite spatial distribution and in 
Subsection \ref{sec:backmod} for the background/foreground objects.
Theoretical and observational justifications for the model choices, along with a schematic illustrating
a possible realization of our model can be found in \paperone.

\subsection{Satellites}
\label{sec:satmod}

\subsubsection{Satellite Spatial Distribution}
\label{subsec:spatDist}
We construct our spatial distribution model to investigate three main components. The first
is the slope of the radial profile of the satellite number density. We model the projected satellite radial distribution as a power-law $P(R)\propto R^{\gamma_p}$. 
The second two parameters describe the ellipticity and orientation of the angular distribution of satellites relative to the host light profile. We model the
satellite angular distribution as elliptical with axis ratio $q_{\rm s}$ between its major and minor axes and with an orientation, $|\phi|$  of its major axis relative to the major axis of the host light profile
$|\phi|$ is allowed to vary between 0 and 90 degrees, where 0 indicates alignment with the major axis of the host light profile. 

Given these parameters, the probability of finding a satellite at a position $(R,\theta)$ is given by:

\begin{equation}
\begin{array}{l}
P(R,\theta|\phi,q_{\rm s},\rpower) \propto \\
\\
R^{\rpower}[\cos^2{(\theta - \phi)} + 1/q_{\rm s}^2\sin^2{(\theta - \phi)}]^{\rpower/2}RdRd\theta
\label{eqn:satelliteSpatDist}
\end{array}
\end{equation}
The normalization of the angular part of this distribution is given by a generalized elliptical integral.

For some of the host galaxies such as face-on spirals and very symmetric early-types, the orientation of
the host major axis is not well defined. To account for this in our inference, we infer the \emph{relative} flattening
between satellites and host galaxies.

Defining the ellipticity of the satellite population to be:
\be
\epss = \frac{1-q_{\rm s}^2}{1+q_{\rm s}^2},
\ee
We relate the ellipticity of the host light profile to the ellipticity of the satellite angular distribution via the parameter $A$:
\be
\epss = \frac{\epsh A}{1 + \epsh (A-1)}
\label{eqn:A}
\ee

The parameter $A$ can be understood as follows: when $A$ is zero, satellites are always distributed isotropically around  host galaxies, regardless of how flattened the host light profiles are. When $A$ is one, the satellite population follows an ellipse with the same elongation as the central light profiles, as $A$ goes to infinity, the distribution approaches a straight line.  
In this work, we only examine whether the satellite angular distribution is rounder or flatter than the host light profile, 
so we restrict our inference to values of $A$ between zero and two.

\subsubsection{Number of Satellites per Host}
To ensure completeness, we study the satellite population brighter than a fixed magnitude contrast from the host galaxies $m_{\rm obj}-m_{\rm{h}}<\dm$, where
the host galaxies in a $\dm$ bin are all at least $\dm$ brighter than the background.  
We model the number of satellites $\numsat$ in a given $\dm$ bin, between $\Rmin<\Rsat<\Rmax$ as constant for all hosts in that bin. 

Both $\Rmin$ and $\Rmax$ are chosen based on observational considerations. $\Rmin$ is the limit where accurate satellite detection is possible given host light subtraction, as described in Section \ref{sec:subtraction}. $\Rmax$ is the limit where a satellite signal is apparent above the background in Figure \ref{fig:prs}, which is 45 $\Rhost$ for early and late-type galaxies. These outer limits correspond on average to 0.5 and 1.0 $\rtwo$ for early and late-type galaxies. We use these average values for $\Rmax$ when performing the inference in units of $\rtwo$.

\subsection{Background/Foreground Objects}
\label{sec:backmod}
We model the background/foreground number density signal around each host as a homogeneous, isotropic signal, with mean surface density $\densb$ with magnitudes brighter than the field magnitude limits ($\mi<25.0$). We inform our inference of $\densb$ by measuring the local background around each of our hosts and using the mean and standard deviation of this distribution to create a prior. 

We follow the method recommended by \citet{Chen++06} of estimating the local background around our host galaxies (rather than taking the field average) in order to accurately measure $\densb$. This is important for removing correlated line of sight structure which is not within the virial radii of the host galaxies. 
In \paperone, we measured the background between 45 and 70 $\Rhost$, where no significant satellite signal was apparent. For consistency, in this analysis, we estimate the background in the same region when scaling distances by $\Rhost$. When performing the analysis in units of $\rtwo$, we calculate the background between 1.0 and 1.5 $\rtwo$, following the simulation results from \citet{Liu++11}.

Various studies have shown that the local projected number density of objects is correlated with host stellar mass and morphology. To account for this, we measure the mean and standard deviations separately for each data set that we run the inference on and use this local background information as a prior when inferring the properties of the satellite distribution.

 For a given value of $\dm$ and host galaxy magnitude, $m_{\rm h}$, only a fraction of the total number of background/foreground objects will appear in our analysis. To calculate this fraction, we model the cumulative distribution function (CDF) of the background number counts by a power-law \citep[e.g.][]{Benitez++04}. Defining $\Sigma_{\rm b,o}$ as the background density measured for all objects with magnitudes brighter than the limiting survey magnitude $m_{\rm{lim}}$, the number density of background/foreground objects around the $j^{\rm th}$ host is given by:
\be
\Sigma_{\rm b, j} = \Sigma_{\rm b,o}
10^{\alpha_{\rm b}(m_{\rm{h}, j} + \dm - m_{\rm lim})}.
\ee
We measure the background slope to be $0.305 \pm 0.005$ near low mass early and late-type galaxies and $0.300 \pm 0.005$ near high mass host galaxies.

\begin{deluxetable*}{clc}[h!]
\tabletypesize{\small}
\tablecaption{\label{tab:modPars}
Summary of model parameters}
\startdata
\hline
Parameter        & Description & Prior \\
\hline
Satellite Model           &             & \\
Ns                        &   Number of satellites per host & U(0,20) \tablenotemark{a}\\
$\rpower$                & Logarithmic slope of the satellite radial distribution & U(-10,0)\\
$A$                      & Relative flattening between satellite angular dist. and host light  & U(0,2) \\
$|\phi|$                 & Angle between the major axis of satellite angular dist. and that of the host light. & U(0,$\pi/2$)\\
\hline
Background Model &        & \\
$\Sigma_{\rm b,o}$       & Number density of all background objects with $\mi<25$ & varies with host mass and $\dm$ \\
$\alpha_b$              & Logarithmic slope of the background number counts & N(0.300,0.005)/N(0.305,0.005)\tablenotetext{b}\\
\enddata
\tablenotetext{a}{U(a,b) denotes a uniform distribution between a and b.}
\tablenotetext{b}{N($\mu$,$\sigma$) denotes a normal distribution with mean $\mu$ and standard deviation $\sigma$.} 
\end{deluxetable*}

\section{Theoretical Prediction for the Number of Satellites per Host}
\label{sec:theornum}
Our choice to model the number of satellites as being constant for all hosts within a fixed magnitude contrast from the host magnitude is a simplification of a more complicated picture.  
Dark matter only simulations predict that dark matter halos should host an approximately constant number of dark matter subhalos with a given \emph{dark matter mass fraction} of the host halo, regardless of the host halo mass \citep[][and references therein]{Krav++10}. 
This scale-invariance relates in a non-trivial way to our chosen observable which is the number of satellites we expect to see within a fixed magnitude contrast from from host halo. This is due to the non-linearity of the stellar-mass to halo-mass relationship for hosts with $\MstarHost > 10^{10} M\odot $ as can be seen, for example, in \citet{Behroozi++10}. 

In recent work, \citet{Busha++11} (hereafter B11)  modeled the number of satellites per $\dm$ m bin as a function of host luminosity using SubHalo Abundance Matching (SHAM) techniques \citep{Kravtsov++04,Behroozi++10}, 
to connect dark-matter halos in the Bolshoi simulations, \citep{Klypin++11,Trujillo++11}  to the r-band luminosity function of galaxies. In order to properly mimic the observational selection function in this work, we have reproduced the measurements of B11 using the luminosity function from the AGES simulations (Cool et al 2012) as applied to the Bolshoi halo catalog at the appropriate redshift. 

To test how the inference method responded to the non-Poissonian distribution of satellites predicted by theory we ran simulations using the distribution of host luminosities in our sample to generate the number of satellites per host based on Equation 8 from B11.  Spatial positions for satellites and background/foreground objects were drawn stochastically from the model described in Section \ref{sec:model}. As desired, our inference accurately returned the mean number of satellites per host along with the other input parameters in the simulations. 

\section{Analysis}
\label{sec:analysis}

 In \paperone, we provide details for the construction of the posterior
probability distribution function (PDF) which allows us to infer
values for the parameters of our model given the data and prior
knowledge of the background listed in ~Table~\ref{tab:modPars}.  For
each parameter in each subset of the data listed in Tables \ref{tab:modResultsEarly} 
and \ref{tab:modResultsLate}, we compute the
posterior PDF using a Markov Chain Monte Carlo (MCMC) method. At least
$10^4$ iterations per chain are performed in order to ensure
convergence.

To study variation with host properties, satellites are analyzed in
bins of `high' and `low' host stellar mass corresponding to
$10.5<\log_{10}[M^{*}_{\rm h}/M\sun]<11.0$, $11.0<\log_{10}[M^{*}_{\rm
h}/M\sun]<11.5$, as well as `low' and `high' host redshift corresponding to
$0.1<z<0.4$ and $0.4<z<0.8$ and early and late-type host galaxies.
Every bin in $\dm$ is analyzed separately for each bin in host morphology, redshift and
stellar mass, using an appropriate prior on
$\Sigma_{\rm b,o}$ estimated using the local background for that
subset of host galaxies.

In order to combine results from different data sets, we bin the
posterior PDFs for the parameters of interest and multiply them
together.  When data sets were analyzed in different regions around
the host, we first re-normalize the inferred satellite numbers using the
posterior median values of $\rpower$ and $\numsat$ to account for the
differences in examined areas, before combining the posteriors.

We tested our inference by running simulations with varying
background and satellite properties. We find that in order to
accurately infer the radial profile $\rpower$ we need a minimum of 50
host galaxies \emph{and} at least 20 satellites.  The first
requirement guarantees that the inference has an adequate estimate of
the background density, and the second that there is a sufficient
satellite signal. Inferring the angular distribution of
satellites is more difficult as this requires two dimensions of
information. We find that at least 50 satellites are necessary to
recover $A$ and $|\phi|$. Furthermore, the inference on $|\phi|$
becomes inaccurate for values of $A$ less than 0.5, thus we do not
report confidence intervals of $|\phi|$ where the posterior favors values of $A$
less than 0.5.

\section{Results}
\label{sec:results}
We divide this section into the three characteristics of the satellite population that our inference constrains, namely the radial and angular distribution of satellites as well as the cumulative luminosity function. For each characteristic we discuss variation with host morphology redshift and stellar mass as well as with satellite luminosity by comparing results in high and low redshift and stellar mass bins as well as in bins of host morphology, as discussed in Section \ref{sec:analysis}.  
Summaries of the inference results for each subsample Tables \ref{tab:modResultsEarly} and \ref{tab:modResultsLate} in the Appendix. To show typical results, Appendix \ref{app:fullPDFs} contains the full bivariate posterior PDFs for all model parameters for high mass, early and late-type hosts for the $\dm = 4.0$ bin.

\subsection{Radial Distribution}
\label{sec:results_gammas}
\begin{figure*}[h!]
\centering
\includegraphics[scale = 0.35, trim = 0 0 20 200, clip =true]{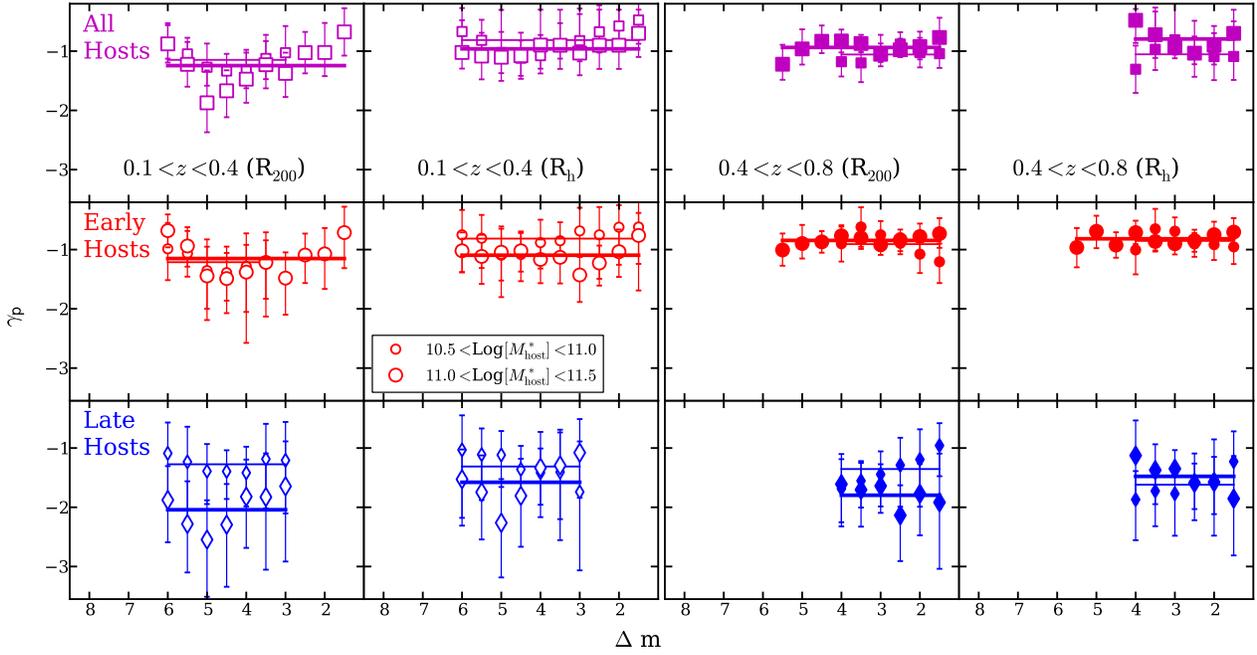}
\caption{Projected radial profiles of the satellite number density distribution with distances scaled by $\Rhost$ and $\rtwo$ for satellites divided by host morphology, redshift and stellar mass. Horizontal lines indicate the average median value of $\rpower$. Large markers indicate results for satellites of higher stellar mass hosts and small markers for the satellites of lower stellar mass hosts.}
\label{fig:gammas}
\end{figure*}

In Figure \ref{fig:gammas} we show the inferred value of the projected slope of the radial profile of the satellite number density as a function of $\dm$, divided by host morphology, stellar mass and redshift. We further compare these results when distances are scaled by $\Rhost$  and by $\Rvir$. 

This figure contains several important results. The first is that there is a significant detection of a population of objects with a power-law radial distribution up to 6.5 magnitudes fainter than low redshift hosts, and 5.5 magnitudes fainter than high redshift hosts. 
Using the average luminosity of the host galaxies in each of those data sets, this corresponds to approximate absolute $r$ band magnitudes of -16.1 and -17.4, respectively. The former is about one magnitude brighter than Sagittarius, the latter is similar to the absolute magnitude of the present day SMC.

Second, there is no significant difference in the inference of $\rpower$ when distances are scaled by $\Rhost$ or by $\Rvir$. This is remarkable given the number of assumptions that go into estimating the virial mass given the stellar mass, and is a very useful result for comparisons between theoretical and observational work for future satellite surveys where accurate host stellar masses are not available.
For simplicity, for the remainder of the paper we discuss results only for distances scaled by $\Rvir$. Results for both distance scalings are listed in Tables are listed in Tables \ref{tab:modResultsEarly} and \ref{tab:modResultsLate} in the Appendix. 

Third, there is no significant variation in $\rpower$ as $\dm$ varies with fixed host properties over ranges as large as 5.5 magnitudes.
This is consistent with the physical processes governing the satellite luminosity function being not spatially related to the host galaxy over this fairly large luminosity range within the distance scales we study, given our measurement uncertainties. 

The average values of $\rpower$ for host galaxies of all morphologies are: $-1.1 \pm 0.1$, $-1.3 \pm 0.4$, $-1.2 \pm 0.1$ and $-1.0 \pm 0.1$ for satellites of low mass-low redshift hosts, high mass-low redshift hosts, low mass-high redshift hosts and high mass-high redshift hosts respectively.  

Individual subsets of host morphology do not show any significant deviation from these values.\footnote{Although the high mass low redshift sample of late-type galaxies appears to have a steeper satellite number density radial profile, this sample is relatively small (only 54 hosts) and does not show a statistically significant deviation.}  As there are no significant differences in the results for these samples, we argue that the satellite population can be well described by a power law with $\rpower =  -1.1 \pm 0.3$ within the level of precision afforded by our data.

\subsection{Angular Distribution}
\label{sec:results_angles}

\begin{figure}[h!]
\centering
\includegraphics[scale = 0.45,trim = 0 0 0 100,clip =true]{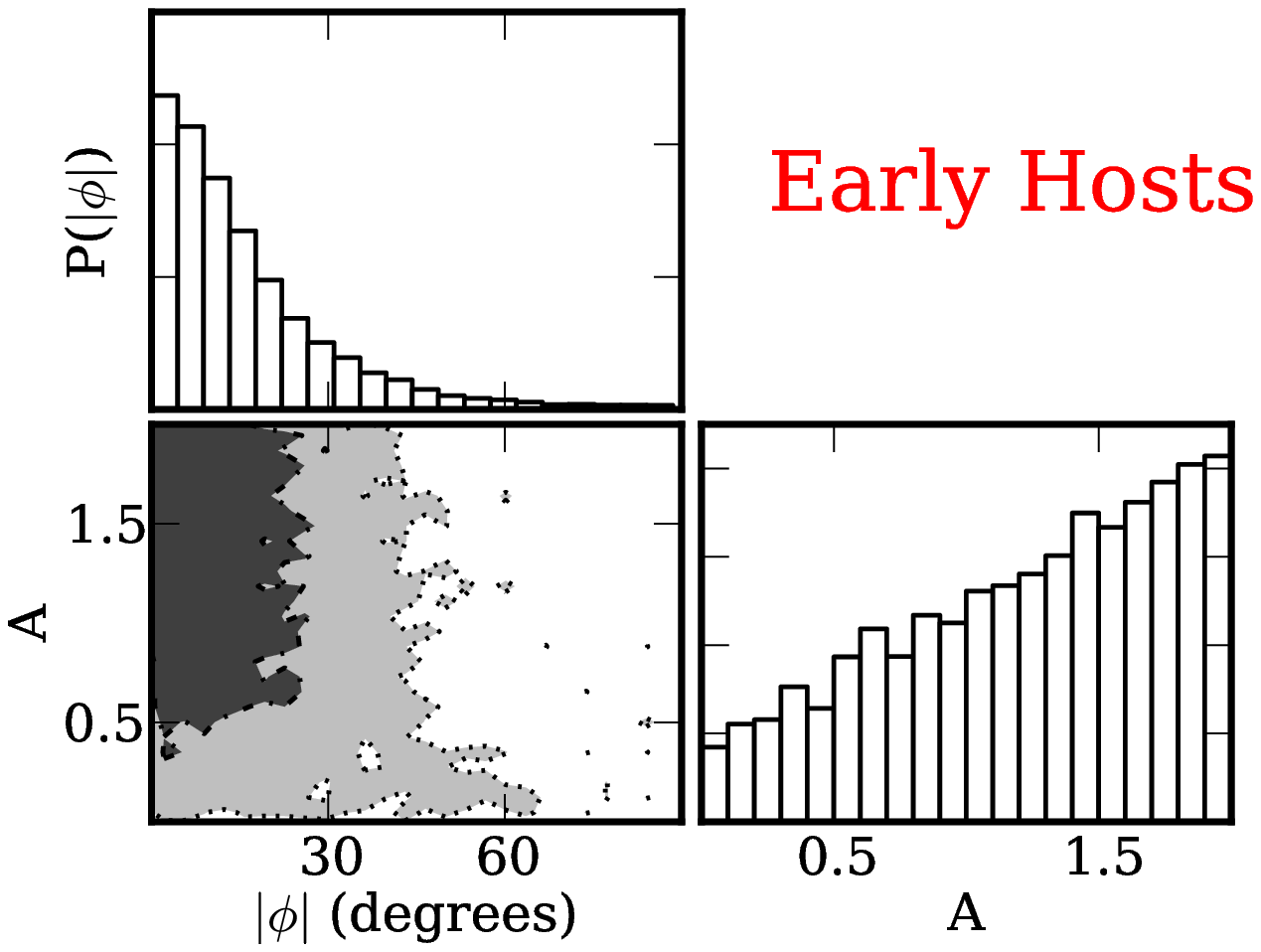}\\
\includegraphics[scale = 0.45, trim = 0 0 0 100,clip =true]{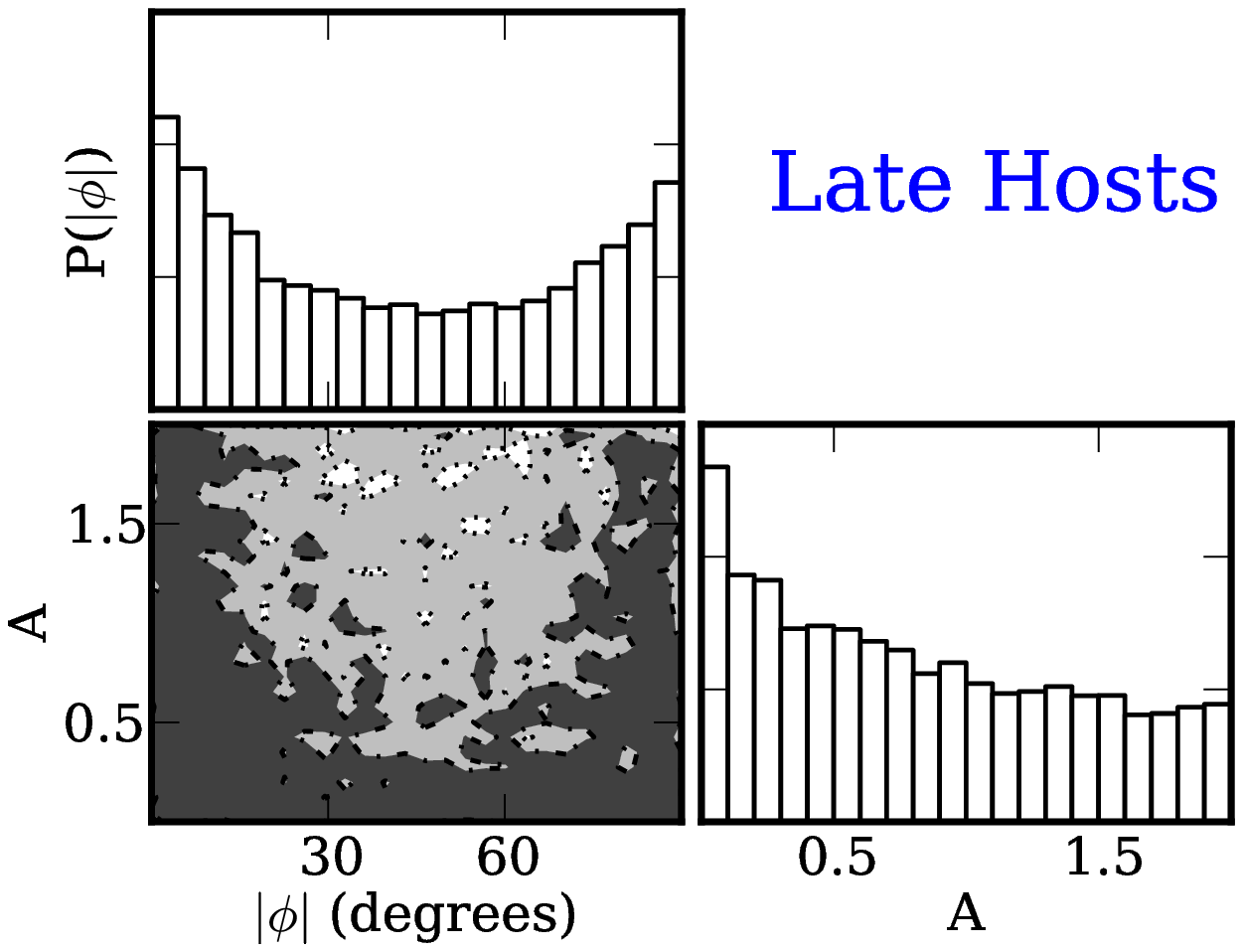}
\caption{Bivariate posterior distributions of $|\phi|$ and $A$ for the satellites of \emph{Upper:} early and \emph{Lower}: late-type host galaxies, for hosts with $0.4<z<0.8$ and $11.0<\log[M_{\rm{host}}^{*}/M\odot]<11.5$}
\label{fig:thetas}
\end{figure}

The satellites of early and late-type galaxies display markedly different angular distributions where the sample of satellites and host galaxies is large enough to allow an inference on the parameters on the angular distribution. As an example, in Figure \ref{fig:thetas} we show the two dimensional bivariate posterior distributions for the parameters $A$ and $|\phi|$ for satellites with $\dm<4$ for early and late-type hosts with $0.4<z<0.8$ and $11.0<$Log[M$^{*}$]$<11.5$. The results for this subsample which are representative of all non-uniform posterior PDFs.
 One sided confidence intervals for $A$ and $|\phi|$ for other non-uniform subsamples are listed in Tables \ref{tab:modResultsEarly} and \ref{tab:modResultsLate}.
Satellites of early-type galaxies show strong anisotropy, with the most likely value of $A$ at 2 and $A>0.9$ at a 68\% confidence level. This indicates that the angular distribution of satellites is more flattened than the light profiles of the early-type galaxies. Furthermore, the satellites of early-type galaxies, are found preferentially along the major axis of the host light profiles. The offset, $|\phi|$ has a most likely value of 0 and is less than 20 degrees at a 68\% confidence level.

In contrast, the average angular distribution of the satellites of late-type galaxies is consistent with being isotropic relative to the host disks, with $A$ having a most likely value of 0 and $A<1.1$ at a 68\% confidence level.  As discussed in Section \ref{sec:analysis}, the inference on $|\phi|$ is unreliable for small values of $A$ as $|\phi|$ has decreasing influence on the likelihood function for small values of $A$. 

There is a concern with regards to describing the ellipticity of the angular distribution of satellites as a fraction, $A$, of the host ellipticity (see Equation \ref{eqn:A}). Namely, late-type galaxies are much more elongated than early-type galaxies but there is no reason to expect the satellites to reflect this. It may be that the satellites of late-type galaxies have anisotropic distributions on average but that the flattening of their distribution is less than the extreme flattening of the disks, making them look isotropic in comparison. To test whether the inferred isotropy is due to the extreme flattening of late-type hosts, we re-ran the inference on late-type hosts, this time artificially changing any late-type axis ratio $b/a$ to 0.6 that was previously flatter than 0.6. We found that the posterior PDF still favored $A=0$.  However, the inference on the angular distribution for late-type hosts is difficult due to the large region obscured by spiral arms. To try to maximize the signal to noise, we performed a combined inference on late-type hosts of all redshifts and stellar masses. In this case, there was evidence for slight anisotropy, with satellites aligned with the galactic disk. Thus, an analysis of the satellites of late-type hosts with a large data set is warranted before any strong conclusions are reached. 

\subsection{Cumulative Luminosity Function}
\label{sec:results_clf}
We detect a significant population of satellites as faint as 6.5 magnitudes fainter than their host galaxies. Results for the inferred number of satellites per host as a function of maximum contrast between host and satellite magnitude ($m_{\rm{host}}-m_{\rm{sat}}<\Delta_m$) can be found in Tables \ref{tab:modResultsEarly} and \ref{tab:modResultsLate}.
A dominant uncertainty in this analysis is caused by the covariance between $\rpower$ and $\numsat$ as can be seen in Figure \ref{fig:pdfsEarly}; the same total number of objects can be achieved if there are more satellites following a shallower radial profile or if there are fewer satellites with a very steep radial profile, and more background objects. We alleviate this degeneracy by performing the inference a second time, using a Gaussian prior on $\rpower$ from the results in Section \ref{sec:results_gammas}. Given that $\rpower$ showed no significant variation given measurement uncertainties as a function of satellite magnitude, host redshift or host morphology (see Section \ref{sec:results_gammas}) we apply the same Gaussian prior on $\rpower$ to all data sets, with mean $-1.1$ and standard deviation of 0.3. 

Results for the inferred number of satellites using the prior are consistent with results without the prior and are listed in Tables \ref{tab:modResultsEarly} and \ref{tab:modResultsLate}
The prior on $\gamma_p$ allows for a significant detection of satellites up to 8 magnitudes, or more than a thousand times fainter than their host galaxies for the low reshift host sample.  Using the average absolute $r$ band magnitude of hosts, this corresponds to satellites with $M_r \sim -13.5/-14.7$ for low and high mass host samples respectively, and is similar to the absolute magnitude of Fornax. The range of $\dm$ does not change for the higher redshift sample but the average measurement uncertainty is decreased.

\begin{figure*}[h!]
\centering
\includegraphics[scale=0.5, trim = 0 0 300 0,clip =true]{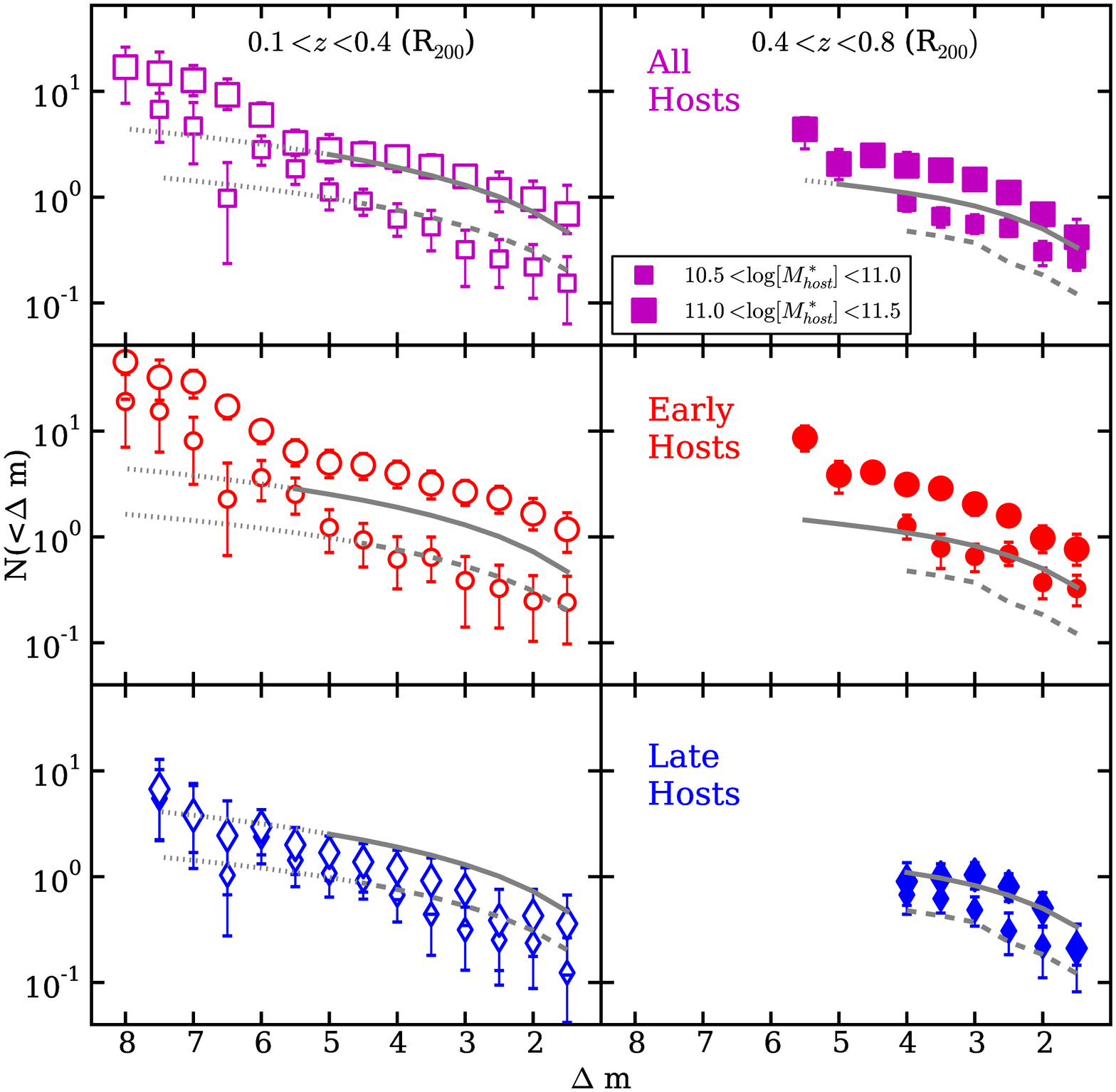}\\
\caption{The cumulative number of satellites per host between 0.07 and 1 $\Rvir$ as a function of the magnitude contrast between host and satellite galaxies, plotted for different samples of host redshift, stellar mass and morphology. Purple squares, red circles and blue diamonds represent all, early and late-type hosts 
respectively. The gray solid and dashed lines are theoretical predictions for the satellites of high and low mass host galaxy samples respectively. Thin dotted lines indicate an extrapolation of the theoretical prediction which was made for satellites brighter than $M_r<-17$. Note that the mean host stellar mass and redshift within each bin shifts slightly towards higher masses and lower redshifts starting at $\dm = 6$ (see Tables \ref{tab:modResultsEarly} and \ref{tab:modResultsLate}).
.}
\label{fig:nsEvol_fixedG}
\end{figure*}

 In Figure \ref{fig:nsEvol_fixedG} we plot the number of satellites per host, $\numsat$, in increasing bins of $\dm$, for varying host morphologies, stellar masses and redshifts between 0.07 and 1.0 $\Rvir$ ($\sim 17-200$ kpc).  The points for `All' morphology hosts come from binning and multiplying the posterior PDFs of the early and late-type hosts \footnote{This is valid as these data sets are independent. The product of the two posterior PDFs can be viewed as a measurement of the number of satellites per host for the entire sample}. We always combine the posteriors even when the inference on the late-type host galaxies is not sufficiently determined to plot (i.e. for $\dm>4$ for more massive, high redshift late-type hosts). The satellite CLF is a fairly constant power-law with slope approximately $N(L)\propto L^{-0.5\pm 0.1}$. There is a slight upward shift in the normalization for satellites with $\dm>6.5$, which is attributable in part to the slight increase in average host stellar masses for these bins (see Tables \ref{tab:modResultsEarly} and \ref{tab:modResultsLate}). We leave a more detailed unbinned analysis of the satellite luminosity function to a future paper.

There is a strong dependence between early-type host stellar mass and
number of satellites within a fixed value of $\dm$.  The dependence on
stellar mass is less apparent for late-type host galaxies, with the
more massive hosts having barely more satellites on average than less
massive host galaxies despite the fact that the low and high mass
late-type hosts have the same average stellar masses as the low and
high mass early-type hosts. While there is a large difference between
the number of satellites of the more massive early and late-type
hosts, the numbers are the same for less massive hosts.

Figure \ref{fig:nsEvol_fixedG} also shows the B11 theoretical prediction for
the mean number of satellites given the distribution of
host luminosities in each sub-sample. The dotted line represents an extrapolation of the theoretical model, to luminosities where the simulation used begins to suffer from incompleteness effects. The theoretical model is in excellent agreement with observation for objects with $0.1 < z < 0.4$, while it tends to underpredict the abundance of satellites of higher redshift objects (although the slopes remains in good agreement). It should be noted, however, that the luminosity function used in the model in the range $0.4 < z < 0.8$ suffers from significant observational uncertainties. Such uncertainties directly impact halo occupation in non-trivial ways, since parameters such as $M_*$ and $\phi_*$ have a significant impart on the mass-luminosity relation for massive objects. Additionally, we assumed a fixed scatter in the mass-luminosity relation of 0.16 dex, which has been shown to be in good agreement with z = 0.1 galaxies (Behroozi et al 2010), but has yet to be explored at higher redshift. A full exploration of the theoretical uncertainties relating to this prediction is beyond the scope of the current paper and is left for future studies.

As expected, subhalo abundance
matching predictions which are constructed without
taking host morphology into account Ð cannot capture
the morphological dependencies. A possible explanation
for the variations is given by the morphology-density relation
 \citep{Dressler++1980, Postman++1984,Tre++03},
 in the sense that at fixed luminosity, early-types
tend to reside in denser environments and have more massive dark matter halos
 than late-types. This in turn might affect the properties of luminous satellites 
and points to limitations of the present SHAM models.
  
\section{Comparison with previous work}
\label{sec:comparison}
We can make several comparisons between our work and low-redshift studies of the satellite population. To start, our inference on $\rpower = -1.1 \pm 0.3$  is consistent with results from the low redshift SDSS study of LRGs by \citet{Watson++10} and of satellites with luminosities similar to that of the LMC \citep{Tollerud++11}. It is slightly shallower, even though marginally consistent, than what was measured by \citet{Chen++08}. Given uncertainties in the scale radii of host galaxies, we are unable to determine whether our result is consistent with NFW as was observed by \citet{Guo++12}. Our result is consistent with our previous work (\paperone), in which we studied the satellites of early-type galaxies from the GOODS fields.

In contrast to \citet{Watson++11}, we do not find a significant trend between $\dm$ and $\rpower$ over the 5 magnitudes in $\dm$ that we studied.
While \citet{Watson++11} measured the satellite radial profile changing from $\rpower \sim -2$ for satellites with $M_r = -21$ to $\rpower \sim -1$ for satellites with $M_r = -18.5$. The closest matching data-set in this study is the low-redshift, high-mass sample, for which we infer $\rpower= -0.7^{+0.4}_{-0.6}$ for $\dm = 1.5$ satellites (which is approximately $M_r\sim -21$) and $-1.7^{+0.6}_{-0.9}$ for $\dm = 4.5$ satellites (approximately $M_r\sim -18.5$). 
 The difference may be due to the fact that our enhanced detection technique allows us to detect a significant new population of faint satellites which would have otherwise remained obscured, causing the radial profile to appear to be flatter. Given our measurement uncertainties we cannot rule out a small increase in the concentration of fainter satellites towards the center as was observed by \citet{Tal++11} and \citet{Guo++12}.

The dependence of the satellite angular distribution on host morphology is consistent with numerous previous low-redshift studies of bright satellites \citep{Bailin++08,Brainerd++05, Agustsson++10}, which also found that satellites of early-type hosts tend to be found along the major axis of the host light profiles.  Furthermore, these works found no significant anisotropy in the angular distribution for the satellites of late-type hosts \citep[see also][]{Yegorova++11}. 

\begin{figure}[h!]
\centering
\includegraphics[scale = 0.35]{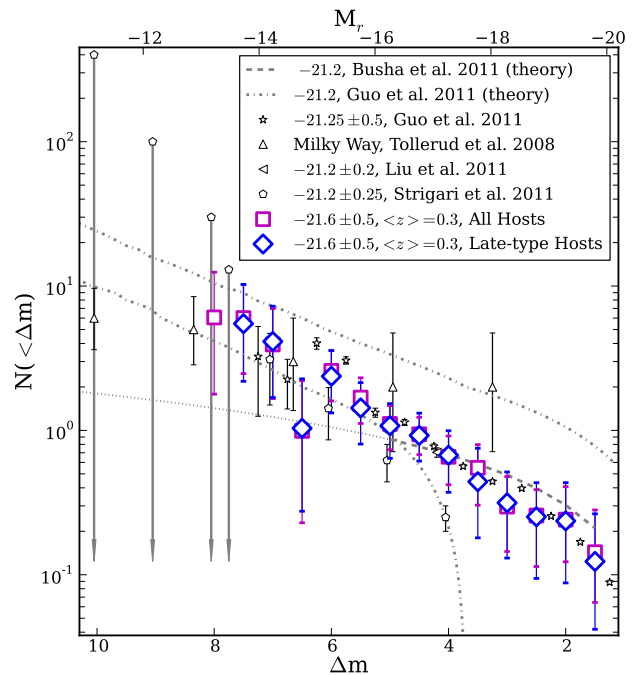}
\caption{Summary of measurements of the satellite population of Milky Way-luminosity hosts. The legend lists the mean value and variation in the r band absolute magnitude for each study. The theoretical prediction from \citet{GuoQi++11} represents the 10 and 90 percentiles of the satellite probability distribution.  Low-mass, low-redshift hosts in this work have a mean redshift of 0.3 and $<M_r>\sim -21.6$ which is brighter than the present day Milky Way. However, accounting for passive evolution of roughly one magnitude from redshift one to present, these galaxies represent approximate Milky Way analogs at redshift 0.3. The upper x-axis is shifted to the left by 0.4 mags for these hosts.}
 \label{fig:compareLFs}
\end{figure}

The satellite numbers we measure for lower redshift hosts are consistent with the results from low redshift SDSS studies of hosts with similar masses. In Figure \ref{fig:compareLFs}, we compare the number of satellites of low mass, low redshift host galaxies to the number in the Milky Way and near other low mass, low redshift host galaxies by \citet{Guo++11}, \citet{Liu++11}, and \citet{Strigari++11}.  The Milky Way has an absolute magnitude of $M_{r} \sim -21.2$, which is  slightly lower than the typical host luminosity in the low stellar mass, low redshift data set in this sample ($M_{r} \sim -21.6$). However we expect some passive evolution in the stellar mass to light ratio of order one between redshifts 0.8 and 0.1 \citep[e.g.][]{Treu++05b}, which would make the average luminosity approximately -21.3 at present day. Furthermore, the typical stellar mass in the low mass, low redshift subset of galaxies is $\log_{10}[M^*_{h}/M\odot]\sim 10.7$ which is equivalent to the stellar mass of the Milky Way within measurement uncertainties \citep{McMillan++11}. Thus the hosts are approximate Milky Way analogs at redshift 0.3. 

In Figure \ref{fig:compareLFs}, we also plot two theoretical predictions. The B11 subhalo abundance matching is identical to that plotted in the corresponding panel in Figure \ref{fig:nsEvol_fixedG}. The second shows the 10-90 percent tails of the predicted satellite distribution from \citet{GuoQi++11} semi-analytic models of galaxy formation applied to the Millennium and Millennium II simulations \citep{Springel++05, Boylan-Kolchin++09}.

At all comparable magnitudes our inferred number of satellites is consistent with that measured by \citet{Guo++11}, \citet{Liu++11} and \citet{Strigari++11}. At the bright end, there are on average fewer satellites per host than what is observed in the Milky Way. The infrequent presence of Magellanic Cloud-equivalent satellites ($\dm= 2.0$) of Milky Way mass hosts has been noted in numerous surveys before this \citep{Guo++11, Lares++11, Liu++11}.  We are able to measure the satellite number down to 8 magnitudes fainter than the host galaxies and corresponds to an average $r$ band luminosity of about -13.3 which is similar to the Fornax satellite. At the faint magnitudes the satellite numbers fall just under the upper limits from \citet{Strigari++11} and the slope of the luminosity function is consistent with that at brighter satellite magnitudes.

At higher redshifts, \citet{Newman++11} measured the pair fraction of galaxies $\dm<2.5$ between redshifts 0.4 and 2. Scaled to the same region as our study, they found $\numsat =0.48\pm 0.09$. This number is somewhat higher than our measurement for low-mass host galaxies which had $0.14\pm 0.1$ in the same region. This discrepancy is likely due to the fact that the authors used global rather than local background subtraction, meaning they estimated the average background across the entire field. This leads to a lower estimate of the number density of background objects than local background estimation as it does not take into account the filamentary clustering of galaxies. The authors estimate that using local rather than global background estimation increases their pair fraction by a factor of two relative to local background techniques, which brings the results into agreement (0.24 $\pm$ 0.05 compared to our measurement of 0.14 $\pm$ 0.1). The authors observed no significant evolution in the pair fraction in the redshift interval they studied.

\begin{figure}[h!]
\centering
\includegraphics[scale = 0.35]{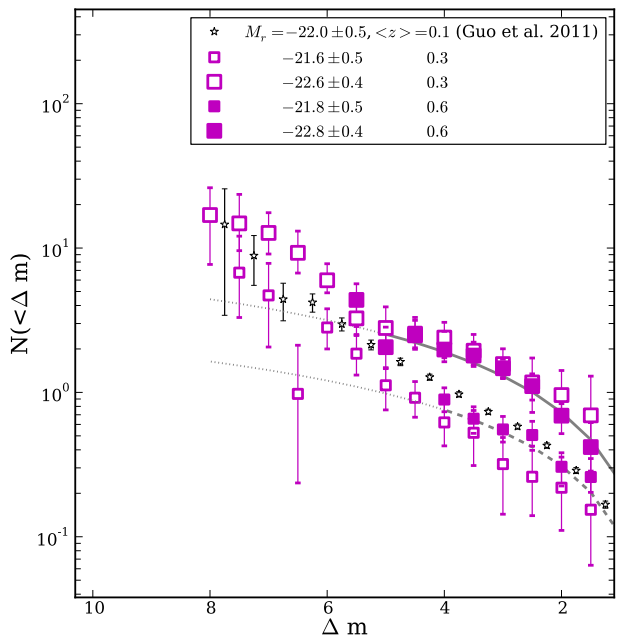}
\caption{Comparison of the cumulative luminosity function of satellites of hosts with varying luminosities. Mean host absolute magnitudes and redshifts are listed in the legend. The gray dashed and solid lines are theoretical predictions for high and low mass, low redshift host galaxies from this work.  The satellite numbers depend strongly on host luminosity,  while there is no significant trend with redshift given measurement uncertainties.}
\label{fig:compareLFs_massive}
\end{figure}

In Figure \ref{fig:compareLFs_massive} we show the satellite CLF for varying host luminosities and redshifts. For comparison, we include the measurement of intermediate luminosity hosts from \citet{Guo++11} at redshift 0.1 along with the theoretical prediction from B11 for redshift 0.3 hosts .  This figure highlights the  the strong dependence of the number of satellites per host on the host stellar mass and luminosity. At the same time, there is no significant redshift evolution in the number of satellites. This highlights the important of selecting constant host mass samples when attempting to study trends in the satellite population.

As shown by the curves, theoretical predictions using the techniques of B11 match
our inferred satellite numbers well for the sample of `All'
morphology hosts at $z = 0.3$, as discussed above.

\section{Discussion}
\label{sec:discussion}
In the Introduction we discussed a large number of physical processes that can be tested by the spatial distribution and number of satellite galaxies. These processes can broadly be divided into three sets. First how satellites are affected by interactions with their host galaxies, second how host galaxies are affected by satellites, and third how our observations fit into the framework of $\Lambda$CDM.  

In terms of the first question, we see no evidence for strong physical interaction between satellite galaxies and central host galaxies on the distance scales we study ($\sim 10-250$ kpc). This is evidenced by the lack of any kind of trend in the steepness of the radial profile of the satellite number density as a function of host redshift, stellar mass, or morphology or as a function of satellite luminosity. In a future paper, we will allow the slope of the radial profile to vary with distance from the host galaxies to allow us to compare the profile with NFW. 
We  see evidence that the satellite angular distribution follows the host mass profile. Assuming that the satellite number density follows the dark matter halo, our result indicates that early-type host galaxies have light profiles that are aligned with their dark matter halos. The alignment for satellites of disk galaxies is weaker indicating that disk galaxies may be less aligned with their dark matter halos. This interpretation of morphologically dependent galaxy-halo alignment is consistent with results from weak lensing \citep{Mandelbaum++06}, and recent theoretical modeling of disk galaxy formation \citep{Deason++11}.  
The fact that satellites of early-type hosts appear in elliptical distributions which are more flattened than the host galaxy light profile may indicate that the extended dark matter halo is more elongated than the galaxy light profile at the center. Several simulations show that baryons can have a dissipational effect on the central dark matter halo, causing it to be more isotropic than it would have been in absence of baryons \citep{Dubinski++94,Kazantzidis++10}. 

With respect to the second question, the constant radial distribution and number of satellites over time indicate a constant accretion of satellites for all host galaxies between redshifts 0.8 and 0.1 for satellites brighter than $\dm<5.5$.
\citet{Newman++11} observed no evolution for bright $\dm<2.4$ pairs between 0.4 and 2.0. The combination of these results indicate that the number of satellites per host has been in equilibrium for at least half the age of the universe. From this we expect that host galaxies have been accreting stellar mass at a fairly constant rate over this time. \citet{Newman++11} estimate that the added stellar mass from the bright pairs is roughly $\sim 6\%$ per Gyr. 

The strong dependence on
host mass in the normalization of the satellites of early-
type galaxies within a fixed bin of $\dm$ reflects the non-linear relationship between host stellar mass and halo
mass. The fact that late-type hosts do not show a significant
corresponding trend may indicate the effects of environment. As discussed above, from the morphology-density
relation, early-type hosts are more likely to be found in
groups, while late-type hosts exist in more isolated environments. This suggests that dark matter halo mass-luminosity relation may also be dependent on environment, something that has been largely ignored in most theoretical models. 
The relationship between host mass and satellite numbers is fundamental to the missing
satellite problem, as clusters of galaxies have satellite
numbers similar to what is predicted by $\Lambda$CDM while isolated field galaxies do not \citep{Krav++10,Klypin++99}. 

Gravitational lensing is currently the only viable method of measuring the halo mass of satellite galaxies outside of the Local Group. \citet{Dalal++02} used flux ratio anomalies in five quadruply-lensed quasars to estimate the fraction of dark matter in satellite galaxies to be between 0.6 and 7 \% near the lensed images, demonstrating the potential of this technique to test the smallest mass scales of $\Lambda$CDM at cosmological distances.  More recently,  \citet{Vegetti++10b, Vegetti++12} used reconstructions of the mass profile of lens galaxies to detect low mass sub-halos near the lensing galaxies. From this \citet{Vegetti++12} estimated a mass fraction of $3.3^{+3.6}_{-1.8}$  in satellites near lensed images. Simulations find a somewhat lower mass fraction in satellites near the Einstein radius \citep{Xu++09} of 0.1\%. 

We can use the results from this work to make an estimate of the fraction of satellites we would expect near massive, intermediate redshift, early-type galaxies (typical of lenses). Assuming a typical Einstein radius of $\sim$ 1", we expect roughly 5 \% of the more massive, early-type host galaxies in our sample to host a satellite with $\dm>7.5$, indicating an average mass fraction in satellite galaxies of a few percent in this region, on average. As discussed in the Introduction, the most massive sub-halos are stripped of dark matter in the inner regions of the host halo. The discrepancy between simulation results and observation may therefore indicate the importance of baryons in the preservation of the most massive satellites close to the central galaxy.

Finally, the detection of significant anisotropy in the satellite angular distribution has important implications for using flux-ratio anomalies to detect satellites. The anisotropy increases the line of sight mass between a lensed image and the host galaxy, effectively making the host galaxy more massive in a particular direction by up to a factor of 6 \citep{Zentner++06}. In a future paper, we will perform a detailed study of the predicted flux ratio anomalies caused by luminous satellites using the results from this work.

\section{Summary}
\label{sec:summary}
We employ the host light subtraction method developed in \paperone~
to study the satellites of 3425 host galaxies selected from
the COSMOS field. The 
depth of the ACS images allows us to measure the satellite luminosity function
more than one thousand times fainter than host galaxies at low redshift, while the combination of
host light subtraction and high resolution images allows us to accurately detect faint sources 
as close as 0.3 (1.4) arcseconds (kpc) to host galaxies.
Using the large volume of the COSMOS field, we examine trends in the
satellite radial distribution, angular distribution and luminosity function. We further examine how
these trends vary with redshift, host stellar mass and host morphology. Our main results
can be summarized as follows:

\begin{enumerate}
  \item The number density of satellites of all host galaxies in the sample
  is well-described by a power-law $P(R)\propto R^{\rpower}$ with $\rpower = -1.1^{+0.3}_{-0.3}$. 
  There is no significant deviation from this as a function of host morphology, stellar mass
  or redshift. Furthermore, there is no evidence for variation in $\rpower$ with satellite luminosity.

  \item The inference results are the same when satellite distances are scaled by the half-width at half-max
  of the host galaxy light profile ($\Rhost$) or by $\Rvir$ determined from the host stellar masses. 
  
  \item Satellites of early-type hosts, follow angular distributions which show strong alignment with the major axis of the host light profile, and fall in an ellipse that is more flattened than the host light profiles. In contrast, the satellites of late-type hosts do not show strong anisotropy. 
  
  \item The satellite cumulative luminosity function (CLF) is a power-law with faint end slope approximately $N(L)\propto L^{-0.5\pm 0.1}$.

  \item The satellite CLF of early-type galaxies shows a strong mass dependence in the normalization, with more massive early-type hosts
  having significantly more satellites than less massive early-type hosts. A similar trend is not apparent for late-type host galaxies. This likely reflects the fact that massive early-type hosts are more likely to be found in groups and exist in more massive dark matter halos than their late-type counterparts. This highlights
  the mass dependence of the satellite luminosity function which is fundamental to the missing satellite problem. 
  
  \item There is no significant evolution evident in the CLFs of satellites between redshift bins of median 0.6 to median 0.3. Satellite numbers
  from the redshift interval 0.1-0.4 are consistent with lower redshift studies ($z<0.1$) of the satellites of host galaxies with equivalent masses.

  \item Predictions from subhalo
abundance matching \citep{Busha++11} broadly
agree with the measured satellite numbers when host morphology is not considered.
However, the theoretical predictions were not created to match trends with host morphology.
We find that such models systematically overpredict the abundance
of satellites surrounding massive late-type hosts, and underpredict the abundance around early-type hosts. As the observations show smaller differences for the satellite abundance around lower mass hosts ($10.5 < \log[M_{h}^{*}/M\odot] < 11.0$), the theoretical model tends to have smaller discrepancies. Additionally, agreement is better at lower redshifts ($z \sim 0.3$) then for higher redshifts ($z \sim 0.6$). As discussed, this is likely due to errors in the model predictions, due to uncertainties in the luminosity function and scatter in the mass-luminosity relation at these higher redshifts.

\end{enumerate}

\acknowledgments
We thank R.~Wechsler, P.~Schneider, D.D.~Xu, L.V.E.~ Koopmans, G.~Dobler and J. Ostriker for many insightful comments and suggestions. We thank O.~Ilbert for providing a stellar mass catalog of objects in the COSMOS field, which greatly enhanced this work. We thank P.~Capak and the rest of the COSMOS team for their work on the COSMOS ACS and ground-based images and catalogs.
AMN and TT acknowledge support
by the NSF through CAREER award NSF-0642621, and by the Packard
Foundation through a Packard Fellowship.  PJM was given support by the
Kavli Foundations and the Royal Society in the form of research fellowships.
Part of the work presented in this paper was performed by AMN, TT, CF, and MB while attending the program "First Galaxies and Faint Dwarfs: Clues to the Small Scale Structure of Cold Dark Matter" at the Kavli Institute of Theoretical Physics at the University of California Santa Barbara.

\bibliographystyle{apj}
\bibliography{references}
\appendix

\input{early_results.tex}

\input{late_results.tex}

\section{Source Extractor Parameters and Photometry Comparison}
\label{app:SE}
After removing host galaxy light from cutout images, we run
\sextractor on the residual image to identify remaining
objects. In this step we use \sextractor parameters tuned to match the
photometry and completeness of the COSMOS photometric catalog. We could not directly the
parameters listed by the COSMOS team, as their photometry was performed on images 
with 0.\farcs03 resolution pixels while at the time of this work, the publicly
available images had 0\farcs05 pixels.  

\begin{figure}[h!]
\centering
\includegraphics[scale=0.4]{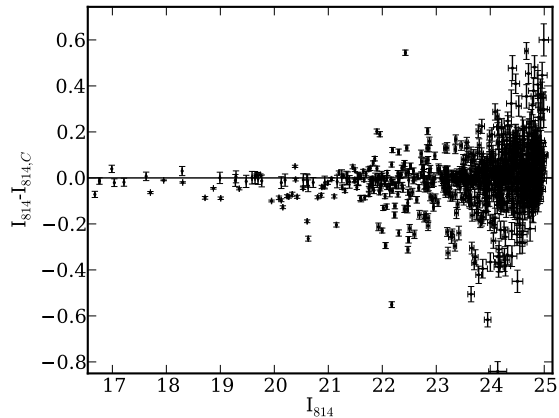}
\caption{Comparison of our photometry with no host galaxy subtraction to COSMOS photometry in an 18 arcmin $^2$ field cutout.}
\label{fig:fieldphot_magauto}
\end{figure}

To check the effects of using different \sextractor parameters and pixel scales, we compare our completeness and photometry to that of the COSMOS photometric catalog in a 4 arcmin$^2$ cutout of the COSMOS field. In this area, we detect 817 objects with I-band MAG\_AUTO$<$25. Of the 817 objects we identified, 792 have matches in the COSMOS photometric catalog to the same depth and within half an arcsecond. In Figure \ref{fig:fieldphot_magauto}, we compare the MAG\_AUTO output from \sextractor to the values in the COSMOS photometric catalog for matching objects. The mean difference in the MAG\_AUTO estimate is $(-2.53 \pm 51)\times 10^{-3}$. The major outliers in the MAG\_AUTO comparison are all in areas of high object density and thus most likely due to differences in deblending.

A full list of our \sextractor parameters is given in Table \ref{table:cos_SePars}
 \begin{deluxetable}{cccc}[h!]
\tabletypesize{\small}
\tablecaption{\label{table:cos_SePars}
\sextractor Parameters}
\renewcommand{\arraystretch}{1.5}
\startdata
\hline\hline   
Parameter        & \multicolumn{2}{c}{Value}                   \\
                 & Large Object Mask & Point Object Mask& Final Photometry     \\
\hline
DETECT\_MINAREA     & 10                & 6              &      6.5     \\
DETECT\_THRESH      & 1.8                & 3              &     1.7   \\
ANALYSIS\_THRESH   & 1.8               & 3             &     1.7   \\
DEBLEND\_NTHRESH & 64                &   64        &      64      \\
DEBLEND\_MINCONT & 1E-6             & 1E-6        &     0.05        \\
FILTER\_NAME          & gauss\_2.5\_5x5.conv    & gauss\_1.5\_3x3.conv            & gauss\_2.5\_5x5.conv \\
BACK\_TYPE             & MANUAL           &                &         \\
BACK\_VALUE           & 0.0                    &                 &  \\
SEEING\_FWHM        &0.1                     &                  &  \\
INTERP\_MAXXLAG	  &   2                    & 	                 & \\
INTERP\_MAXYLAG   &   2 		      &    		& \\
WEIGHT\_TYPE         & MAP\_RMS       & 			 &\\
\enddata
\tablecomments{Parameters not listed are \sextractor default  values.
Blank spaces indicate that the parameter is the same for all rows.}
\end{deluxetable}

\section{Completeness}
\begin{figure*}[h!]
\centering
\includegraphics[scale = 0.4]{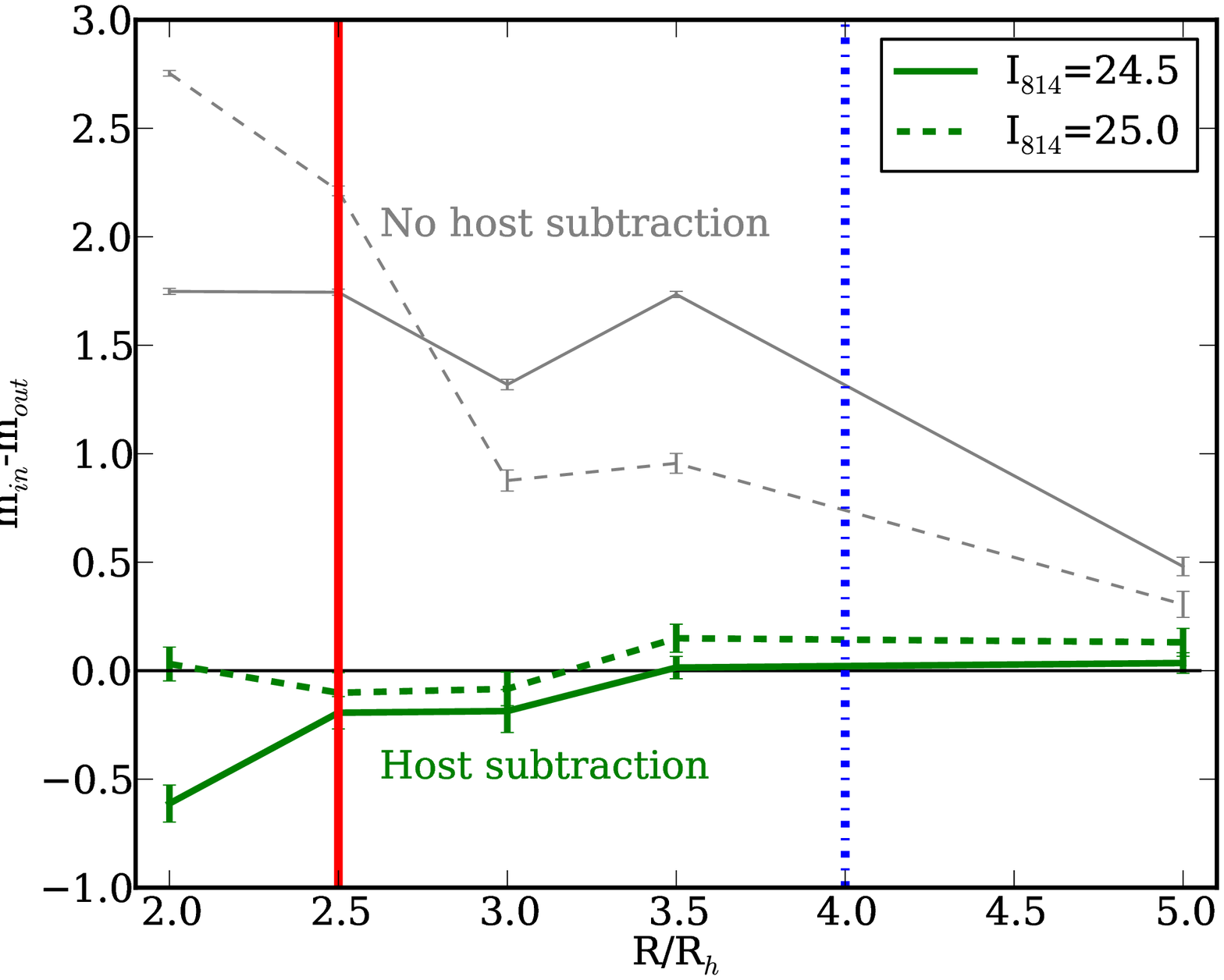}\includegraphics[scale = 0.4]{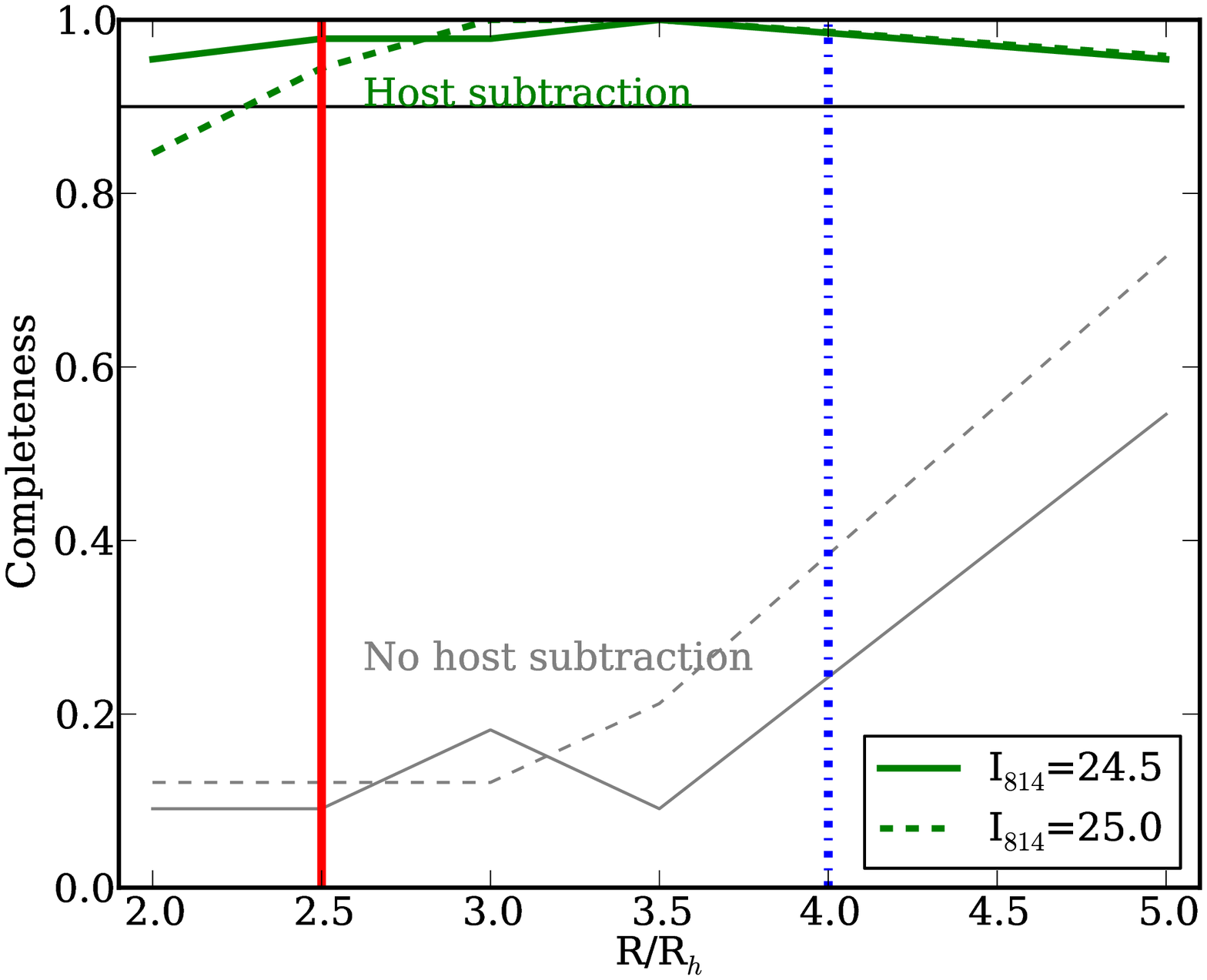}
\caption{Results of completeness simulations. Red solid and blue dot-dashed vertical lines show the imposed minimum radii at which we study the satellite population for early and late-type galaxies respectively. On the left, the horizontal black line indicates that the input and output magnitudes are the same. On the right the horizontal line indicates 90\% completeness. Host subtraction is essential both for photometric accuracy and for completeness at small radii.}
\label{fig:completeness}
\end{figure*}

To estimate our sensitivity to low surface brightness companions to host galaxies, we simulate dwarf galaxy-like objects in the ACS images with Sersic profiles given by \citet{deRijcke++09} at a range of redshifts representative of the redshift range studied in this work. We then perform the host subtraction routine and estimate the fraction of objects we recover and the output magnitude relative to the input magnitude. The results of these simulations are shown in Figure \ref{fig:completeness}. For comparison we also show the results without host subtraction but with object detection performed using the same final \sextractor parameters as are used in the host-subtracted case. 

We choose the minimum radius at which we search for satellites where completeness begins to drop below 90\% for early-type hosts. For late-type hosts, the minimum radius is chosen to ensure spiral arms are excluded from the analysis. Images after host subtraction are checked for every host to ensure that these minimum radii exclude spiral arms or extended residuals from disk structure (see \paperone).

\label{app:completeness}

\section{Full Posterior PDFs}
\label{app:fullPDFs}

\begin{figure*}[h!]
\centering
\includegraphics[scale = 0.32,trim = 20 0 130 130, clip = true]{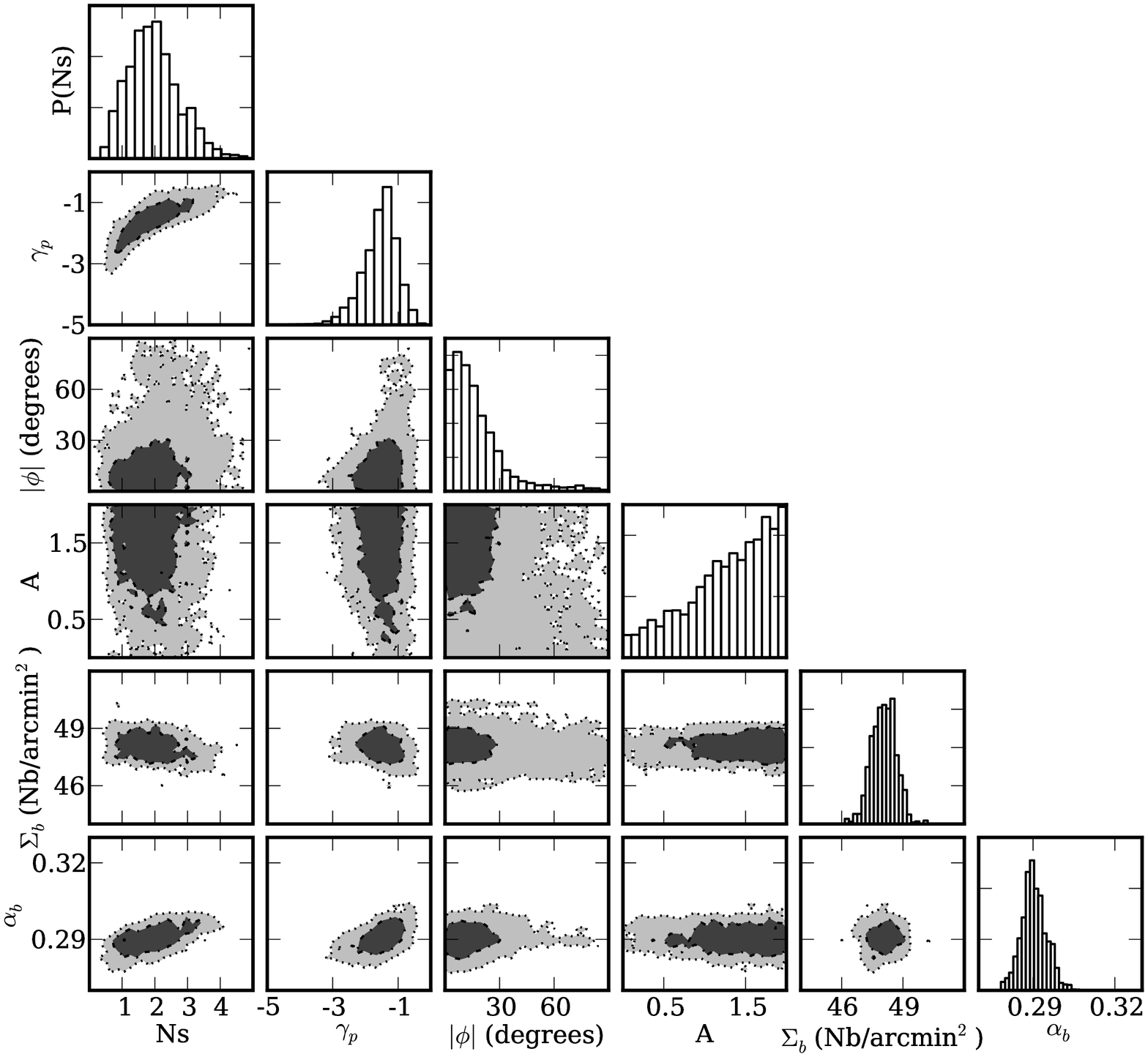}\includegraphics[scale = 0.32, trim = 20 0 130 130, clip=true]{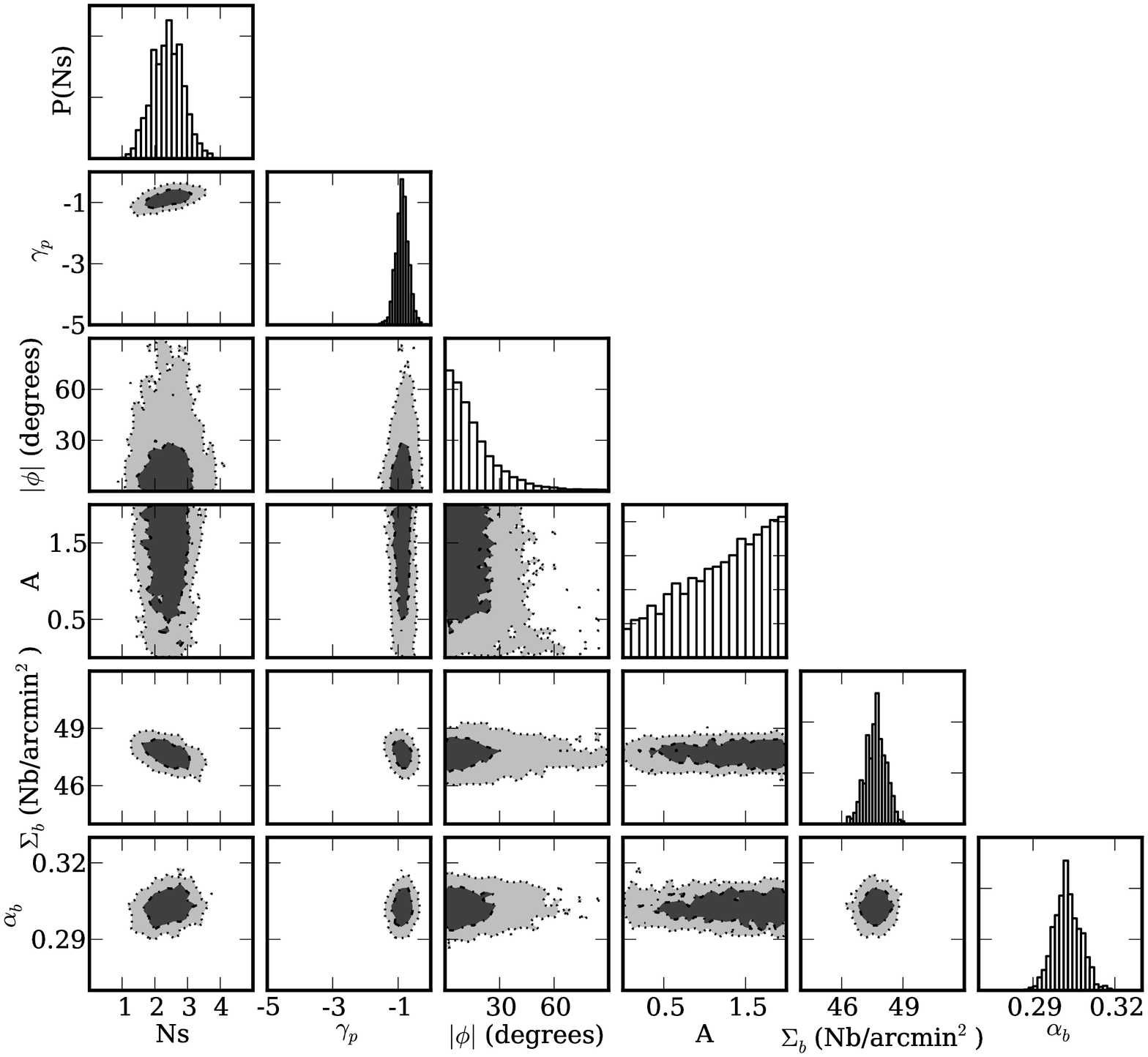}
\includegraphics[scale = 0.32,trim = 20 0 130 90, clip = true]{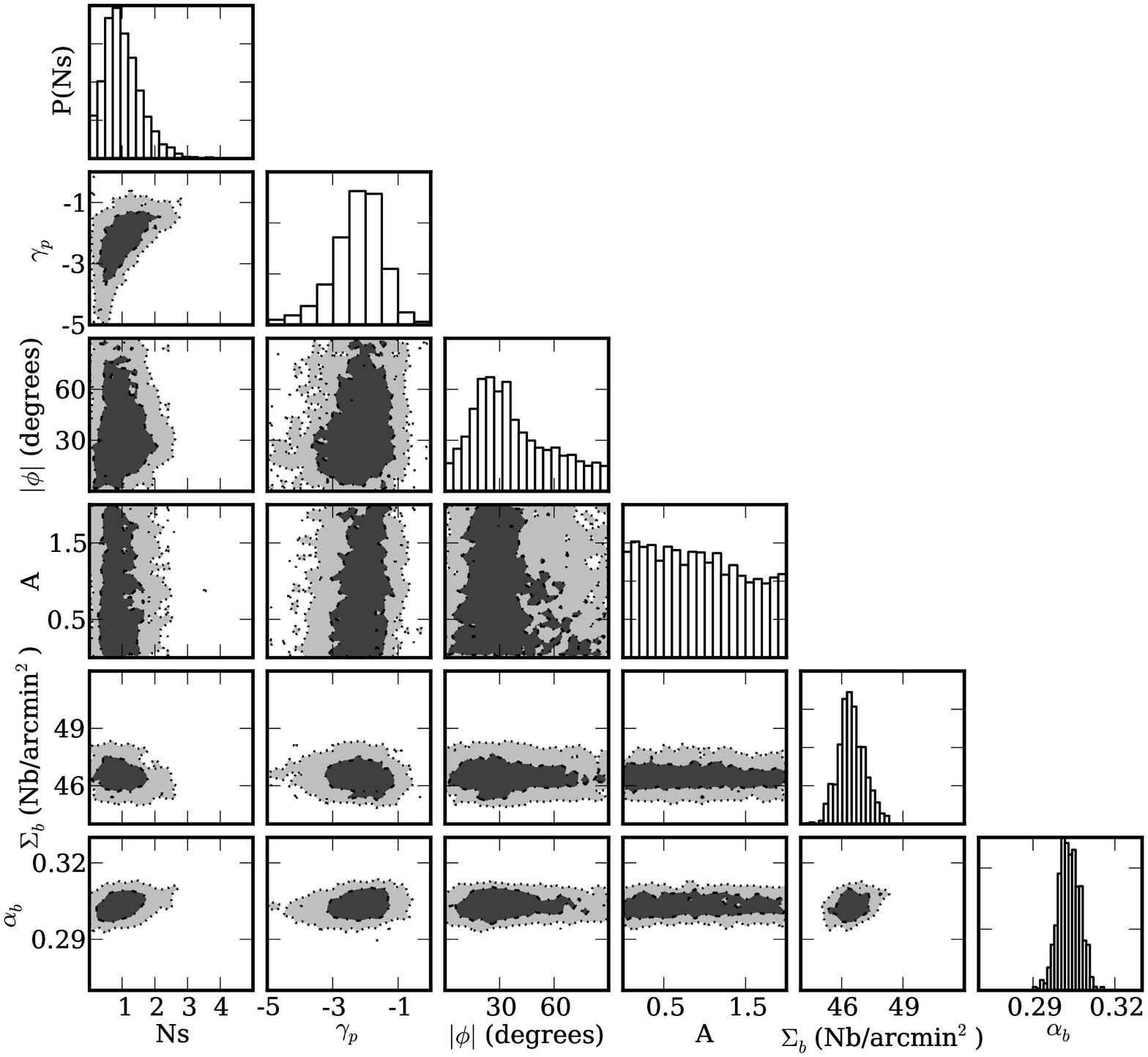}\includegraphics[scale = 0.32,trim = 20 0 130 130, clip=true]{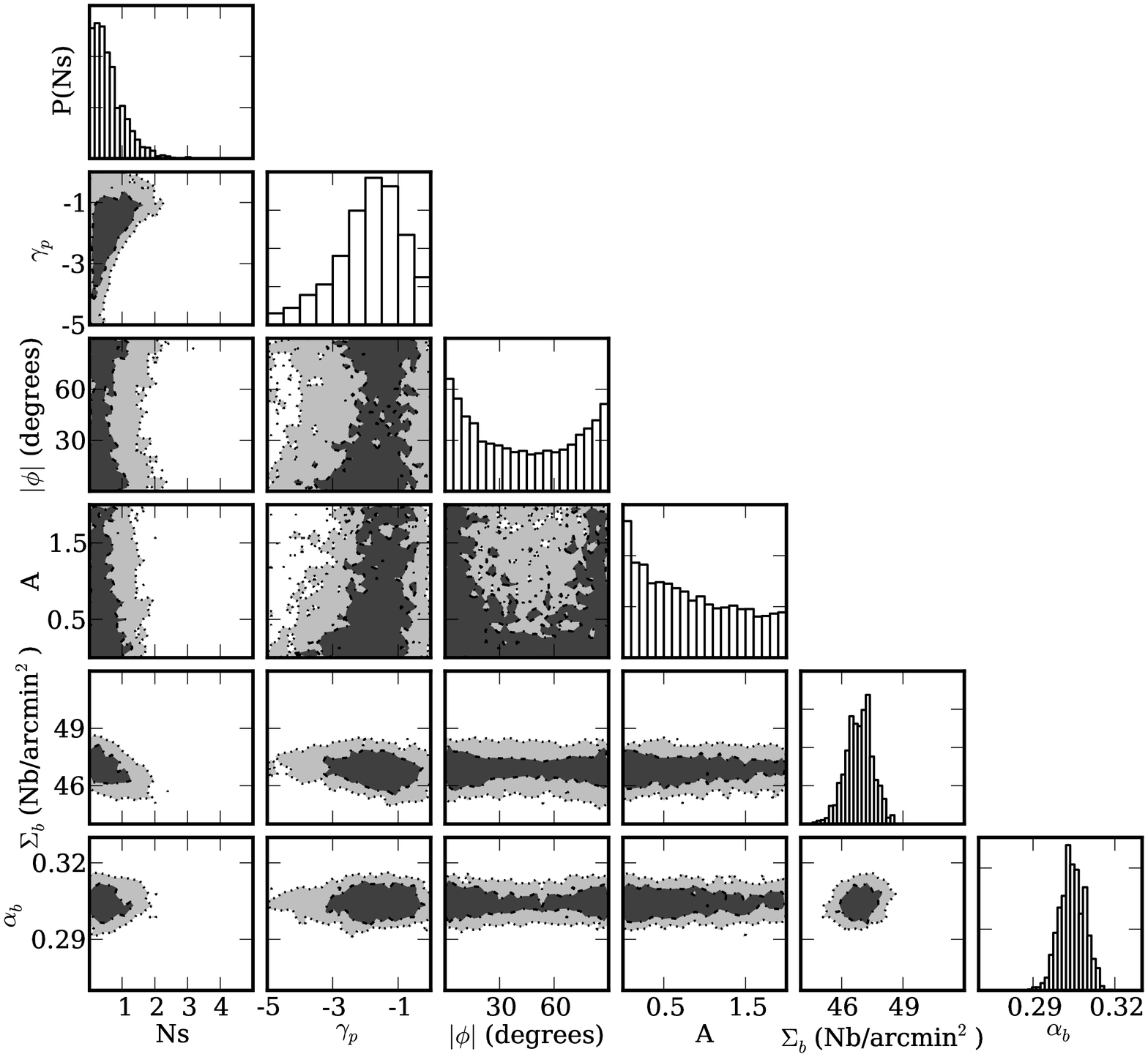}
\caption{Full posterior PDFs for $\dm$ =4.0 satellites of high mass, \emph{clockwise from top left}: Low redshift early-type hosts, high redshift early-type hosts, high redshift late-type hosts, low redshift late-type hosts. Dark and light gray contours indicate regions of 68 and 95\% confidence respectively.}
\label{fig:pdfsEarly}
\end{figure*}

\end{document}

%% file: macros.tex
\def\apj{ApJ}
\def\mnras{MNRAS}
\def\apjl{ApJL}
\def\apjs{ApJS}
\def\araa{ARAA}
\def\aap{A\&A}

\def\paperone{N11}
\def\GC{GOODS/COSMOS}

\def\galfit{{\sc galfit }}
\def\hst{{\it HST }}
\def\sextractor{{\sc SExtractor }}

\def\eg{{\it e.g.\ }}
\def\ie{{\it i.e.\ }}

\def\be{\begin{equation}}
\def\ee{\end{equation}}
\def\zband{$z_{850}$-band }
\def\vband{$v_{606}$-band }
\def\z8{$z_{850}$}
\def\z{z$_{850}$}
\def\I{I$_{814}$}
\def\isub{i$^{+}$}

\newcommand{\bs}{\boldsymbol}
\newcommand{\mc}{\mathcal}
\newcommand{\true}[1]{\widehat{#1}}

\def\TODO#1#2{{\bf TODO: {#1}: {#2}}}
\def\QUERY#1#2{{\bf QUESTION: {#1}: {#2}}}

\def\sun{\odot}
\def\Msun{M_\odot}
\def\ss{SML}
\def\ls{LS}
\def\Mstar{M^{*}}
\def\rtwo{R_{200}}

\def\modpars{\bs{\theta},\bs{\Lambda}}
\def\allDat{\bs{D}}
\def\data{\bs{d}}

\def\spatpars{\bs{\theta}}
\def\epss{\epsilon_{\rm s}}
\def\rpower{\gamma_{\rm p}}
\def\qfrac{f_{\rm q}}
\def\pars{\bs{\theta}_{\rm s}}
\def\phispiral{\phi_{\rm sp}}
\def\phiel{\phi_{\rm el}}
\def\dtheta{\phi}
\def\pos{\bs{x}}
\def\Rsat{R_{\rm sat}}
\def\Rmin{R_{\rm min}}
\def\Rmax{R_{\rm max}}

\def\alphas{\alpha_{\rm s}}
\def\Lsat{L_{\rm s}}
\def\lamo{\lambda_{\rm o}}
\def\lam{\lambda}
\def\mi{I_{814}} 
\def\mz{z_{850}} 
\def\mmax{m_{\rm{max}, j}}
\def\nobs{N^{\rm obs}}
\def\mumax{\mu_{\rm max}}
\def\mumin{\mu_{\rm min}}
\def\nobjs{N}
\def\dm{\Delta m}

\def\numsat{N_{\rm{s} }}

\def\parb{\bs{\theta}_b}
\def\densb{\Sigma_b}
\def\sigdensb{\sigma_{b,cos}}
\def\alphab{\alpha_{\rm b}}
\def\numbak{N_{\rm b}}
\def\numbakobs{N_{\rm b,obs}}
\def\numbakmod{N_{rm b, mod}}

\def\cosvar{\xi}
\def\cosvarnorm{\xi_{\rm o}}
\def\cosvarpower{\delta}
\def\mmax{m_{\rm max}}

\def\epsh{\epsilon_{\rm h}}
\def\hdat{\bs{h}}
\def\numhosts{N_{\rm h}}
\def\Lhost{L_{\rm h}}
\def\Rhost{R_{\rm h}}
\def\Rh{$R_{\rm h}$}
\def\Rvir{R_{200}}
\def\Mvhost{M_{\rm v, h}}
\def\MstarHost{M^{*}_{\rm h}}
\def\Mvir{M_{vir}}

\def\sat{S}
\def\back{B}
\def\prob{{\rm Pr}}
\def\Gauss{{\rm G}}

\def\nobjscorrect{\hat{N}_{\rm c}}

%% file: early_results.tex
\section{Host galaxy properties and inference results}                       
\begin{deluxetable*}{clclc|cccc|ll|c}[h!]
\tabletypesize{\small}
\tablecaption{\label{tab:modResultsEarly}
Summary of early-type host galaxy properties and results from the inference using distances scaled by $\Rvir$}
\tablehead{\colhead{$\dm$} &
           \colhead{N$_{\rm h}$} &
           \colhead{$\log[\Mstar/\Msun]$\tablenotemark{a}}  &               
           \colhead{$z$\tablenotemark{b}}&            
           \colhead{$M_r$\tablenotemark{a}} &
	     \colhead{$\numsat$\tablenotemark{c}} &
	     \colhead{$\rpower$ \tablenotemark{c} } &
	     \colhead{$A$\tablenotemark{c} \tablenotemark{d}} &
	     \colhead{$|\phi|$\tablenotemark{c} \tablenotemark{d}} &
	     \colhead{$\numsat$ \tablenotemark{e}} &
	     \colhead{$\rpower$ \tablenotemark{e}}&
           \colhead{Ns\tablenotemark{c}\tablenotemark{f}} }

\startdata
\cutinhead{$10.5< \log[\Mstar/\Msun$]$<11.0$, 0.1$<$z$<$0.4, $0.02<R/\Rvir<0.5$, $2.5<R/\Rhost<45$} 
1.5 & 210 & $10.7\pm0.1$ & $0.31\pm0.07$& $-21.7\pm0.4$ & \nodata \tablenotemark{g}           &\nodata&\nodata   & \nodata &\nodata&\nodata &$0.13^{+0.1}_{-0.08}$\\
2.0   & 210 &$10.7\pm0.1$ &$0.31\pm0.07$ & $-21.7\pm0.4$ &\nodata & \nodata &\nodata & \nodata& \nodata& \nodata & $0.14^{+0.1}_{-0.08}$\\
2.5 &210  &$10.7\pm0.1$&$0.31\pm0.07$& $-21.7\pm0.4$& \nodata   & \nodata & \nodata  & \nodata & \nodata &\nodata& $0.2^{+0.1}_{-0.1}$\\
3.0   &210 &$10.7\pm0.1$ &$0.31\pm0.07$ &$-21.7\pm0.4$& \nodata&\nodata &  \nodata & \nodata& \nodata & \nodata & $0.2^{+0.1}_{-0.1}$\\
3.5 &210 &$10.7\pm0.1$& $0.31\pm0.07$ & $-21.7\pm0.4$ &$0.3^{+0.2}_{-0.2}$   & $-1.2^{+0.5}_{-0.9}$& \nodata&\nodata&$0.6^{+0.2}_{-0.2}$& $-0.9^{+0.3}_{-0.3}$ & $0.4^{+0.2}_{-0.1}$\\
4.0 & 210 &$10.7\pm0.1$& $0.31\pm0.07$ &$-21.7\pm0.4$ & $0.2^{+0.3}_{-0.2}$  & $-1.3^{+0.6}_{-1}$& $>0.7$   & $<48$ &$0.5^{+0.3}_{-0.3}$& $-0.9^{+0.4}_{-0.4}$&  $0.3^{+0.2}_{-0.2}$\\
4.5 & 208 &$10.7\pm0.1$&$0.31\pm0.07$& $-21.7\pm0.4$& $0.4^{+0.3}_{-0.2}$  & $-1.4^{+0.4}_{-0.5}$& $>0.7$ & $<24$ &$0.6^{+0.3}_{-0.3}$& $-1.1^{+0.4}_{-0.5}$& $0.5^{+0.2}_{-0.2}$     \\
5.0  & 182 &$10.7\pm0.1$& $0.31\pm0.07$ &$-21.7\pm0.4$ &  $0.5^{+0.4}_{-0.3}$ & $-1.4^{+0.5}_{-0.6}$& $>0.9$ & $<26$ &$0.7^{+0.4}_{-0.4}$& $-1.1^{+0.5}_{-0.7}$& $0.7^{+0.3}_{-0.3}$    \\
5.5 & 182 &$10.7\pm0.1$ & $0.31\pm0.07$&   $-21.8\pm0.4$ &  $1.4^{+0.6}_{-0.6}$ & $-1.1^{+0.4}_{-0.4}$& $>0.7$ & $<31$ &$1.6^{+0.8}_{-0.7}$& $-0.8^{+0.4}_{-0.5}$& $1.4^{+0.6}_{-0.5}$   \\
6.0   & 58  &$10.8\pm0.1$ & $0.23\pm0.06$&  $-21.8\pm0.4$ & $2^{+1}_{-1}$        & $-1.0^{+0.4}_{-0.5}$& $>0.7$ & $<51$ &$2^{+1}_{-1}$& $-0.8^{+0.4}_{-0.6}$& $2.0^{+0.9}_{-0.8}$   \\
6.5 & 28  &$10.8\pm0.1$ & $0.19\pm0.05$& $-21.6\pm0.4$ & \nodata &  \nodata &\nodata& \nodata & \nodata &\nodata& $1.3^{+2}_{-0.9}$\\
7.0   & 18  & $10.9\pm0.1$ & $0.18\pm0.05$  &$-21.7\pm0.5$  &\nodata &\nodata & \nodata &\nodata&\nodata &\nodata& $5^{+3}_{-3}$\\
7.5 &  5  & $10.9\pm0.1$& $0.13\pm0.03$  & $-21.5\pm0.4$ &\nodata &\nodata  &\nodata& \nodata&\nodata &\nodata & $9^{+7}_{-5}$\\
8.0 & 3   & $10.9\pm0.1$ &$0.12\pm0.01$ & $-21.5\pm0.3$ & \nodata& \nodata  &\nodata&\nodata& \nodata & \nodata & $11^{+9}_{-7}$\\

\cutinhead{$11.0<\log[\Mstar/\Msun$]$<11.5$, 0.1$<$z$<$0.4, $0.05<R/\Rvir<0.5$, $4<R/\Rhost<45$ }                              
1.5 &92 &11.2$\pm$0.1&  0.32$\pm$0.06  & $-22.6\pm0.4$ & $0.6^{+0.3}_{-0.3}$ & $-0.7^{+0.4}_{-0.6}$& $>0.7$   & $<47$  &$0.5^{+0.3}_{-0.3}$&$-0.8^{+0.4}_{-0.5}$ & $0.6^{+0.3}_{-0.2}$\\
2.0   &92 &11.2$\pm$0.1 &0.32$\pm$0.06  &$-22.6\pm0.4$&$ 0.8^{+0.3}_{-0.3}$   & $-1.1^{+0.4}_{-0.6}$& $>0.7$ & $<36$  &$0.6^{+0.3}_{-0.3}$&$-1.0^{+0.4}_{-0.4}$&$0.9^{+0.3}_{-0.3}$\\
2.5 &92& 11.2$\pm$0.1& 0.32$\pm$0.06 &$-22.6\pm0.4$& $ 1.1^{+0.4}_{-0.4}$   & $-1.1^{+0.4}_{-0.5}$&\nodata & \nodata &$0.9^{+0.3}_{-0.3}$&$-1.2^{+0.3}_{-0.4}$& $1.2^{+0.4}_{-0.3}$\\
3.0   &92 &11.2$\pm$0.1&0.32$\pm$0.06 &$-22.6\pm0.4$& $ 1.1^{+0.5}_{-0.4}$  & $-1.5^{+0.4}_{-0.6}$&\nodata & \nodata &$0.9^{+0.4}_{-0.3}$&$-1.4^{+0.3}_{-0.5}$& $1.4^{+0.4}_{-0.4}$\\ 
3.5 &92& 11.2$\pm$0.1& 0.32$\pm$0.06 &$-22.6\pm0.4$& $ 1.5^{+0.7}_{-0.7}$ & $-1.2^{+0.4}_{-0.6}$& $>0.7$  & $<31$     &$1.2^{+0.6}_{-0.5}$&$-1.1^{+0.3}_{-0.4}$ & $1.6^{+0.5}_{-0.5}$\\  
4.0 &92& 11.2$\pm$0.1&0.32$\pm$0.06  &$-22.6\pm0.4$& $ 1.7^{+0.8}_{-0.7}$  & $-1.4^{+0.5}_{-0.7}$& $>0.9$ & $<20$    &$1.5^{+0.7}_{-0.6}$&$-1.2^{+0.3}_{-0.4}$& $2.1^{+0.6}_{-0.6}$ \\
4.5 &92& 11.2$\pm$0.1& 0.32$\pm$0.06 &$-22.6\pm0.4$& $ 1.9^{+0.9}_{-0.7}$  & $-1.5^{+0.4}_{-0.6}$&  $>1.1$ & $<18$    &$1.7^{+0.7}_{-0.7}$&$-1.0^{+0.3}_{-0.4}$& $2.5^{+0.7}_{-0.7}$ \\  
5.0  &92& 11.2$\pm$0.1& 0.32$\pm$0.06 & $-22.6\pm0.4$ & $ 2.0^{+1.0}_{-0.9}$  & $-1.4^{+0.5}_{-0.7}$&  $>1.0$ & $<25$ &$1.4^{+0.8}_{-0.7}$&$-1.0^{+0.4}_{-0.5}$ & $2.6^{+0.8}_{-0.7}$ \\
5.5 &92 &11.2$\pm$0.1& 0.32$\pm$0.06 &$-22.6\pm0.4$ & $ 3^{+1}_{-1}$        & $-0.9^{+0.3}_{-0.3}$& $>0.7$   & $<66$  &$1.6^{+0.9}_{-0.8}$&$-1.1^{+0.4}_{-0.5}$ & $3.3^{+1.0}_{-0.9}$\\
6.0   & 81 & 11.2$\pm$0.1& 0.31$\pm$0.06  &$-22.6\pm0.4$ & $ 6^{+1}_{-1}$   &  $-0.7^{+0.3}_{-0.3}$&  $>0.7$   & $<67$ &$3^{+1}_{-1}$&$-1.0^{+0.3}_{-0.4}$  & $5^{+1}_{-1}$\\
6.5 & 39 & 11.3$\pm$0.1& 0.27$\pm$0.07  &$-22.6\pm0.4$ & $ 9^{+2}_{-2}$   &  $-0.7^{+0.3}_{-0.3}$&   $>0.7$  & $<58$  &$3^{+2}_{-2}$&$-1.7^{+0.5}_{-0.9}$  & $9^{+2}_{-2}$\\    
7.0   &   15 &11.3$\pm$0.1& 0.21$\pm$0.04 & $-22.6\pm0.4$& \nodata & \nodata &\nodata& \nodata & \nodata& \nodata& $15^{+4}_{-4}$\\ 
7.5  &  9  &11.3$\pm$0.1& 0.19 $\pm$0.04 & $-22.6\pm0.4$ & \nodata & \nodata& \nodata &\nodata&\nodata &\nodata & $17^{+8}_{-7}$\\
8.0 &    3 &11.3$\pm$0.1 &0.19 $\pm$0.04 &$-22.7\pm0.4$&\nodata&  \nodata  & \nodata & \nodata&\nodata &  \nodata & $23^{+16}_{-13}$\\

\cutinhead{$10.5<\log[\Mstar/\Msun] <11.0$, 0.4$<$z$<$0.8, $0.02<R/\Rvir<0.5$, $2.5<R/\Rhost<45$}
1.5 & 1038 & 10.7$\pm$0.1 & 0.6$\pm$0.1 & $-21.8\pm0.5$ &  $0.16^{+0.07}_{-0.06}$ & $-1.2^{+0.3}_{-0.4}$ &  $>0.7$ & $<27$ &$0.17^{+0.06}_{-0.05}$&$-1.0^{+0.3}_{-0.3}$ &  $0.18^{+0.06}_{-0.06}$\\
2.0   & 1019 &10.7$\pm$0.1 &0.6$\pm$0.1 &$-21.8\pm0.5$& $ 0.21^{+0.08}_{-0.08}$   & $-1.1^{+0.3}_{-0.3}$ & \nodata & \nodata & $0.20^{+0.05}_{-0.06}$&$-0.9^{+0.3}_{-0.3}$& $0.21^{+0.08}_{-0.06}$\\     
2.5 & 997  &10.7$\pm$0.1& 0.6$\pm$0.1 & $-21.8\pm0.5$& $ 0.41^{+0.09}_{-0.09}$  &$-0.9^{+0.2}_{-0.2}$ & \nodata&\nodata & $0.30^{+0.07}_{-0.08}$&$-0.9^{+0.2}_{-0.2}$& $0.38^{+0.1}_{-0.08}$ \\
3.0 & 893  &10.7$\pm$0.1& 0.6$\pm$0.1 &$-21.9\pm0.4$   &  $ 0.4^{+0.1}_{-0.1}$ & $-0.8^{+0.2}_{-0.3}$&\nodata & \nodata &$0.37^{+0.1}_{-0.09}$&$-0.7^{+0.2}_{-0.2}$   & $0.4^{+0.1}_{-0.1}$\\
3.5 & 642  & 10.8$\pm$0.1& 0.6$\pm$0.1  &$-22.0\pm0.4$ &  $ 0.5^{+0.2}_{-0.2}$    &$-0.6^{+0.3}_{-0.4}$& \nodata & \nodata &$0.4^{+0.1}_{-0.1}$&$-0.6^{+0.3}_{-0.4}$ & $0.4^{+0.2}_{-0.2}$\\
4.0   &  372 &10.8$\pm$0.1 & 0.5$\pm$0.1 & $-22.0\pm0.4$ &  $0.8^{+0.2}_{-0.2}$     & $-0.9^{+0.3}_{-0.3}$&\nodata&\nodata &$0.5^{+0.2}_{-0.2}$&$-1.0^{+0.3}_{-0.4}$ & $0.7^{+0.2}_{-0.2}$  \\

\cutinhead{$11.0 <\log\Mstar/\Msun]<11.5$, 0.4$<$z$<$0.8, $0.02<R/\Rvir<0.5$, $2.5<R/\Rhost<45$ }         
1.5 &331  & 11.2$\pm$0.1& 0.6$\pm$0.1& $-22.8\pm0.4$ & $ 0.5 ^{+0.1}_{-0.2}$& $ -0.7^{+0.3}_{-0.2}$ & $>0.7$  &$<54$       & $0.5^{+0.1}_{-0.1}$& $-0.7^{+0.2}_{-0.2}$& $0.4^{+0.2}_{-0.1}$\\
2.0   & 331 &11.2$\pm$0.1&0.6$\pm$0.1 &  $-22.8\pm0.4$ & $ 0.6^{+0.2}_{-0.2}$ &  $-0.8^{+0.2}_{-0.3}$ &  \nodata& \nodata & $0.6^{+0.1}_{-0.1}$& $-0.7^{+0.2}_{-0.2}$& $0.5^{+0.2}_{-0.1}$\\
2.5 &331 &11.2$\pm$0.1 &0.6$\pm$0.1&  $-22.8\pm0.4$ & $ 1.0^{+0.2}_{-0.2}$ &  $-0.8^{+0.2}_{-0.2}$ &\nodata& \nodata & $0.8^{+0.2}_{-0.2}$& $-0.8^{+0.2}_{-0.2}$& $0.9^{+0.2}_{-0.2}$ \\
3.0 &331  &11.2$\pm$0.1& 0.6$\pm$0.1 & $-22.8\pm0.4$  & $ 1.2^{+0.2}_{-0.2}$ &  $-0.9^{+0.2}_{-0.2}$ & $>0.8$ & $<16$  & $1.0^{+0.2}_{-0.2}$& $-0.9^{+0.2}_{-0.2}$& $1.1^{+0.3}_{-0.2}$      \\
3.5 & 322 &11.2$\pm$0.1& 0.6$\pm$0.1 & $-22.8\pm0.4$  & $ 1.7^{+0.3}_{-0.3}$ &  $-0.8^{+0.1}_{-0.2}$ & $>0.7$  &$<21$   & $1.2^{+0.3}_{-0.2}$& $-0.9^{+0.2}_{-0.2}$& $1.6^{+0.3}_{-0.3}$   \\
4.0   & 272 &11.2$\pm$0.1&0.6$\pm$0.1 &  $-22.8\pm0.4$  & $ 1.8^{+0.3}_{-0.3}$ &  $-0.8^{+0.2}_{-0.2}$ & $>0.8$ & $<28$  & $1.4^{+0.3}_{-0.3}$& $-0.7^{+0.2}_{-0.2}$& $1.7^{+0.3}_{-0.3}$    \\
4.5 & 170 &11.2$\pm$0.1 & 0.6$\pm$0.1 & $-22.8\pm0.4$ & $ 2.4^{+0.5}_{-0.5}$ &  $-0.9^{+0.2}_{-0.2}$ & $>1.0$ & $<18$  & $2.0^{+0.4}_{-0.4}$& $-0.9^{+0.2}_{-0.2}$& $2.3^{+0.5}_{-0.5}$    \\
5.0   & 86  &11.2$\pm$0.1& 0.5$\pm$0.1 &  $-22.8\pm0.4$ & $ 2.2^{+0.7}_{-0.7}$ &  $-0.9^{+0.3}_{-0.3}$ & $>0.8$ & $<26$  & $2.7^{+0.8}_{-0.8}$& $-0.7^{+0.3}_{-0.3}$& $2.1^{+0.7}_{-0.7}$     \\ 
5.5 & 31  &11.2$\pm$0.1& 0.5$\pm$0.1 & $-22.9\pm0.3$ & $  5^{+2}_{-1}$      &  $-1.0^{+0.3}_{-0.3}$ & $>0.7$  & $<23$  & $4^{+2}_{-1}$& $-1.0^{+0.3}_{-0.3}$& 
$5^{+1}_{-1}$\\
\tablenotetext{a}{Geometric means}
\tablenotetext{b}{Mean}
\tablenotetext{c}{Satellite distances scaled by $\Rvir$}
\tablenotetext{d}{68 \% one-sided confidence interval}
\tablenotetext{e}{Satellite distances scaled by $\Rhost$}
\tablenotetext{f}{Inference performed with a Gaussian prior on $\rpower$ with mean $-1.1$ and standard deviation}
\tablenotetext{g}{Blank spaces indicate no inference on the parameter as the posterior PDF was essentially uniform}
\end{deluxetable*}

%% file: late_results.tex
\begin{deluxetable*}{lllll|llll|ll|l}[h!]
\tabletypesize{\small}
\tablecaption{\label{tab:modResultsLate}
Summary of late-type host galaxy properties and inference results}
\tablehead{\colhead{$\dm$} &
           \colhead{N$_{\rm h}$} &
           \colhead{$\log[\Mstar/\Msun]$\tablenotemark{a}}  &               
           \colhead{$z$\tablenotemark{b}}&            
           \colhead{$M_r$\tablenotemark{a}} &
	     \colhead{$\numsat$\tablenotemark{c}} &
	     \colhead{$\rpower$ \tablenotemark{c} } &
	     \colhead{$A$\tablenotemark{c} \tablenotemark{d}} &
	     \colhead{$|\phi|$\tablenotemark{c} \tablenotemark{d}} &
	     \colhead{$\numsat$ \tablenotemark{e}} &
	     \colhead{$\rpower$ \tablenotemark{e}}&
           \colhead{Ns\tablenotemark{c}\tablenotemark{f}} }
\startdata
\cutinhead{$10.5<\log[\Mstar/\Msun$]$<11.0$,   0.1$<$z$<$0.4, $0.07<R/\Rvir<1.0$, $4<R/\Rhost<45$ } 
1.5 & 274 &	10.7$\pm$0.5	& 0.31$\pm$0.07 &	$-21.6\pm 0.5$& \nodata \tablenotemark{g}   &	\nodata & \nodata  & \nodata    & \nodata   &\nodata  &$0.11^{+0.2}_{-0.08}$\\ 
2.0   &  274    &10.7$\pm$0.5   & 0.31$\pm$0.07 & $-21.6\pm 0.5$   &\nodata 							&\nodata  &\nodata  &  \nodata   & \nodata &\nodata     &$0.2^{+0.2}_{-0.1}$  \\
2.5 &  274& 10.7$\pm$0.5   &   0.31$\pm$0.07  & $-21.6\pm 0.5$   &\nodata    						&\nodata& \nodata     &  \nodata   &\nodata  &    \nodata    & $0.2^{+0.2}_{-0.1}$  \\
3.0   &   274   & 10.7$\pm$0.5&  0.31$\pm$0.07  &$-21.6\pm 0.5$    & $0.2^{+0.3}_{-0.2}$   & $-1.2^{+0.6}_{-0.9}$ & $<1.3$  &\nodata  &$0.2^{+0.2}_{-0.1}$&$-1.7^{+0.9}_{-1}$     &  $0.3^{+0.2}_{-0.1}$  \\
3.5 &   274   & 10.7$\pm$0.5 & 0.31$\pm$0.07  &$-21.6\pm 0.5$   & $0.3^{+0.3}_{-0.2}$   & $-1.2^{+0.6}_{-0.8}$ & $<1.3$  & \nodata  &$0.2^{+0.3}_{-0.2}$&$-1.4^{+0.7}_{-1}$	 & $0.4^{+0.3}_{-0.2}$   \\
4.0 &   274   & 10.7$\pm$0.5  & 0.31$\pm$0.07 &$-21.6\pm 0.5$   & $0.5^{+0.3}_{-0.3}$   & $-1.4^{+0.4}_{-0.6}$ & $<1.3$  &\nodata     &$0.6^{+0.3}_{-0.3}$&$-1.4^{+0.4}_{-0.5}$  & $0.6^{+0.3}_{-0.3}$    \\
4.5 & 262 & 10.7$\pm$0.5  &   0.31$\pm$0.07    & $-21.6\pm 0.5$   & $0.8^{+0.4}_{-0.3}$   & $-1.4^{+0.4}_{-0.5}$ &\nodata &\nodata &$0.8^{+0.3}_{-0.4}$&$-1.4^{+0.4}_{-0.5}$ & $0.8^{+0.4}_{-0.3}$    \\
5.0   & 228  & 10.7$\pm$0.5 & 0.30$\pm$0.07     &$-21.7\pm 0.5$ & $0.8^{+0.5}_{-0.4}$   & $-1.4^{+0.5}_{-0.6}$ &$<1.3$   & \nodata &$1.0^{+0.5}_{-0.5}$&$-1.1^{+0.4}_{-0.5}	$ &   $1.0^{+0.4}_{-0.4}$   \\
5.5 & 158  & 10.7$\pm$0.5 & 0.28$\pm$0.08     &$-21.7\pm 0.5$ & $1.1^{+1}_{-0.7}$     & $-1.2^{+0.6}_{-1}$ & $<1.3$  & \nodata    	 &$1.4^{+0.8}_{-0.8}$&$-1.1^{+0.4}_{-0.8}$	& $1.2^{+0.7}_{-0.5}$   \\
6.0   &  94  & 10.8$\pm$0.1 & 0.25$\pm$0.08     & $-21.8 \pm 0.5$& $2^{+2}_{-1}$             & $-1.1^{+0.5}_{-0.9}$   & $<1.3$  &\nodata     &$2^{+1}_{-1}$&$-1.0^{+0.6}_{-1}$&  $2^{+1}_{-1}$   \\
6.5 & 39  &  10.7$\pm$0.1 & 0.18$\pm$0.05     &$-21.6 \pm 0.5$ &\nodata                     &\nodata                          &  \nodata &      \nodata &    \nodata &	\nodata 		                   & $0.9^{+1}_{-0.7}$   \\
7.0   & 25  &  10.7$\pm$0.1 & 0.16$\pm$0.04     & $-21.6 \pm 0.5$ &\nodata                    &\nodata 						& \nodata  &  \nodata    &	\nodata    &	\nodata 						&                                         $3^{+3}_{-2}$   \\
7.5 & 15  &  10.7$\pm$0.1 & 0.13$\pm$0.03     &$-21.4\pm0.4$ &\nodata 				&\nodata 						& 	\nodata &    \nodata	&	\nodata 	&\nodata                                    & $4^{+4}_{-3}$   \\

\cutinhead{$11.0<\log[\Mstar/\Msun]<11.5$,   0.1$<$z$<$0.4, $0.07<R/\Rvir<1.0$, $4<R/\Rhost<45$ }
1.5 & 54  &11.2$\pm$0.1& 0.32 $\pm$0.06  & $-22.3 \pm 0.4$       &\nodata                      &\nodata                   & \nodata  & \nodata     &  \nodata    				& \nodata   &$0.3^{+0.3}_{-0.2}$   \\           
2.0   & 54    &  11.2$\pm$0.1 & 0.32 $\pm$0.06  &  $-22.3 \pm 0.4$  &\nodata                        &      \nodata        & \nodata   & \nodata      &  \nodata  					& \nodata     & $0.3^{+0.3}_{-0.2}$  \\
2.5 & 54    & 11.2$\pm$0.1 &0.32 $\pm$0.06   & $-22.3 \pm 0.4$   &\nodata                        &   \nodata              & \nodata    & \nodata     &  \nodata   			 &\nodata & $0.3^{+0.3}_{-0.2}$   \\
3.0   &  54   &  11.2$\pm$0.1 & 0.32 $\pm$0.06  &$-22.3 \pm 0.4$    & $  0.5^{+0.5}_{-0.3}$ & $-1.6^{+0.8}_{-1}$ & \nodata & \nodata &$ 0.8^{+0.6}_{-0.5}$   & $-1.1^{+0.6}_{-0.8}$ &$0.5^{+0.5}_{-0.3}$   \\
3.5 & 54    & 11.2$\pm$0.1   & 0.32 $\pm$0.06   & $-22.3 \pm 0.4$ & $  0.6^{+0.5}_{-0.4}$ & $-1.8^{+0.7}_{-1}$   &\nodata & \nodata   & $0.8^{+0.6}_{-0.5}$ & $-1.3^{+0.6}_{-0.9}$ & $0.7^{+0.5}_{-0.4}$    \\
4.0 &  54   & 11.2$\pm$0.1   & 0.32 $\pm$0.06   &$-22.3 \pm 0.4$ & $  0.8^{+0.6}_{-0.5}$ & $-1.8^{+0.6}_{-0.9}$ & \nodata & \nodata  &$ 1.0^{+0.7}_{-0.6}$ & $-1.3^{+0.6}_{-0.8}$ & $0.9^{+0.6}_{-0.4}$     \\
4.5 &  54   & 11.2$\pm$0.1   & 0.32 $\pm$0.06   & $-22.3 \pm 0.4$ & $  0.8^{+0.6}_{-0.5}$ & $-2.3^{+0.7}_{-1}$ &\nodata   & \nodata  & $1.2^{+0.8}_{-0.6}$ & $-1.8^{+0.6}_{-0.9}$ & $1.0^{+0.5}_{-0.4}$     \\
5.0  &   54  &  11.2$\pm$0.1   &  0.32 $\pm$0.06   &$-22.3 \pm 0.4$& $  1.0^{+0.6}_{-0.5}$ & $-2.5^{+0.7}_{-1}$ &  \nodata &\nodata  & $1.0^{+0.6}_{-0.5}$ & $-2.3^{+0.7}_{-0.9}$ & $1.2^{+0.6}_{-0.5}$  \\
5.5 &  54  &   11.2$\pm$0.1   &0.32 $\pm$0.06   &  $-22.3 \pm 0.4$ & $  1.2^{+0.8}_{-0.5}$ & $-2.3^{+0.7}_{-0.8}$   &  \nodata   &\nodata     & $1.4^{+1}_{-0.7}$ & $-1.7^{+0.6}_{-0.8}$ & $1.4^{+0.8}_{-0.6}$   \\
6.0   & 45   &   11.2$\pm$0.1  & 0.31$\pm$0.06   & $-22.4 \pm 0.4$   & $  2^{+1}_{-1}$          & $-1.9^{+0.6}_{-0.7}$ & \nodata   & \nodata      & $2^{+2}_{-1}$      & $-1.5^{+0.5}_{-0.8}$& $2^{+1}_{-1}$       \\
6.5 & 14   &  11.2$\pm$0.1  & 0.24$\pm$0.04   & $-22.3 \pm 0.5$     &\nodata                      &	\nodata 			  &    \nodata      &	 \nodata   &  \nodata            &\nodata                     & $2^{+2}_{-1}$       \\
7.0   & 9  &     11.2$\pm$0.1  & 0.23$\pm$0.04   & $-22.4\pm 0.5 $      &\nodata                      &	\nodata 			  &  \nodata        &	 \nodata     &    \nodata         &\nodata                    & $3^{+3}_{-2}$      \\
7.5 &	9   &   11.2$\pm$0.1    &	0.23$\pm$0.04 & $-22.4\pm 0.5 $  &\nodata 		         &	\nodata 		      &  \nodata         &\nodata 	       &   \nodata         &\nodata                     & $7^{+5}_{-4}$      \\

\cutinhead{$10.5<\log[\Mstar/\Msun$]$<11.0$, 0.4$<$z$<$0.8, $0.07<R/\Rvir<1.0$, $4<R/\Rhost<45$}
1.5  &	857 & 10.7 $\pm$0.1    & 0.6$\pm$0.1   & $-21.8 \pm 0.5$  &  $  0.2^{+0.2}_{-0.1}$ & $-1.0^{+0.4}_{-0.5}$ & $<1.0$ & \nodata     &   $ 0.13^{+0.08}_{-0.08}$ & $-1.2^{+0.5}_{-0.6}$ &  $0.2^{+0.1}_{-0.1}$ \\ 
2.0 &  857   & 10.7 $\pm$0.1&  0.6$\pm$0.1 & $-21.8 \pm 0.5$  &  $  0.2^{+0.1}_{-0.1}$ & $-1.2^{+0.5}_{-0.8}$ & $<1.1$ &  \nodata    &$  0.09^{+0.08}_{-0.07}$          & $-1.5^{+0.7}_{-0.9}$     &  $0.19^{+0.1}_{-0.09}$\\
2.5 & 814 &10.7 $\pm$0.1    &0.6$\pm$0.1& $-21.8 \pm 0.5$       &   $  0.2^{+0.2}_{-0.1}$ & $-1.3^{+0.5}_{-0.7}$ & $<1.1$ & \nodata     &$  0.16^{+0.1}_{-0.07}$     & $-1.6^{+0.5}_{-0.6}$   &  $0.3^{+0.2}_{-0.1}$\\
3.0 & 726 & 10.7 $\pm$0.1   &0.6$\pm$0.1 & $-21.9 \pm 0.5$      &  $  0.4^{+0.2}_{-0.1}$ & $-1.4^{+0.4}_{-0.5}$ & $<1.1$ & \nodata     &$  0.2^{+0.1}_{-0.1}$       & $-1.8^{+0.5}_{-0.7}$       &  $0.4^{+0.2}_{-0.1}$\\
3.5 & 583 &10.7 $\pm$0.1    & 0.6$\pm$0.1& $-22.0 \pm 0.5$       &  $  0.5^{+0.2}_{-0.2}$ & $-1.6^{+0.3}_{-0.5}$ & $<0.7$ &\nodata      &$  0.3^{+0.2}_{-0.2}$      & $-1.7^{+0.4}_{-0.6}$     &  $0.5^{+0.2}_{-0.2}$ \\
4.0 & 372 & 10.7 $\pm$0.1 & 0.5 $\pm$0.1 & $-22.1 \pm 0.5$      &  $  0.5^{+0.3}_{-0.2}$ & $-1.7^{+0.5}_{-0.6}$ & $<1.1$ & \nodata     &$  0.4^{+0.2}_{-0.2}$      & $-1.9^{+0.5}_{-0.7}$     &  $0.5^{+0.3}_{-0.2}$\\

\cutinhead{$11.0<\log[\Mstar/\Msun$]$<11.5$, 0.4$<$z$<$0.8, $0.07<R/\Rvir<1.0$, $4<R/\Rhost<45$} 
1.5 & 182 & 11.2$\pm$0.1 & 0.6$\pm$0.1  & $-22.5 \pm 0.5$   & $  0.1^{+0.1}_{-0.09}$ & $-2.0^{+0.8}_{-1}$   &$<1.2$  & \nodata       &$ 0.16^{+0.1}_{-0.09}$  & $ -1.9^{+0.7}_{-1}$ &$0.2^{+0.1}_{-0.1}$       \\ 
2.0   &  182    &   11.2$\pm$0.1  &  0.6$\pm$0.1 &$-22.5 \pm 0.5$    & $  0.3^{+0.2}_{-0.2}$ & $-1.8^{+0.5}_{-0.7}$ &$<1.2$  & \nodata    &$  0.4^{+0.2}_{-0.2}$  & $  -1.6^{+0.4}_{-0.5}$ &$0.4^{+0.2}_{-0.1}$        \\
2.5 & 182     &   11.2$\pm$0.1   & 0.6$\pm$0.1 & $-22.5 \pm 0.5$   & $ 0.5^{+0.2}_{-0.2}$ & $-2.1^{+0.5}_{-0.8}$ &$<0.8$&\nodata   &$0.7^{+0.3}_{-0.2}$  & $  -1.6^{+0.3}_{-0.4}$ & $0.6^{+0.2}_{-0.2}$ \\
3.0 & 181 & 11.2$\pm$0.1  & 0.6$\pm$0.1 & $-22.5 \pm 0.5$  & $  0.9^{+0.3}_{-0.3}$ & $-1.6^{+0.4}_{-0.5}$ & $<0.9$ & \nodata    &$  1.0^{+0.3}_{-0.3}$  & $  -1.3^{+0.3}_{-0.4}$ &$0.8^{+0.3}_{-0.3}$ \\
3.5 & 170 & 11.2$\pm$0.1  & 0.6$\pm$0.1  &$-22.6 \pm 0.4$  & $  0.7^{+0.4}_{-0.3}$ & $-1.7^{+0.5}_{-0.6}$ & $<1.0$ & \nodata   &$  0.8^{+0.4}_{-0.3}$  & $  -1.4^{+0.4}_{-0.6}$ &  $0.7^{+0.3}_{-0.2}$ \\ 
4.0   & 139 & 11.2$\pm$0.1    & 0.6$\pm$0.1 &$-22.6 \pm 0.4$   & $  0.7^{+0.5}_{-0.4}$ & $-1.6^{+0.5}_{-0.7}$ & $<1.1$ &\nodata  &$  0.8^{+0.6}_{-0.5}$& $  -1.1^{+0.6}_{-0.8}$ &$0.7^{+0.4}_{-0.3}$ \\

\tablenotetext{a}{Geometric means}
\tablenotetext{b}{Mean}
\tablenotetext{c}{Satellite distances scaled by $\Rvir$}
\tablenotetext{d}{68 \% one-sided confidence interval}
\tablenotetext{e}{Satellite distances scaled by $\Rhost$}
\tablenotetext{f}{Inference performed with a Gaussian prior on $\rpower$ with mean $-1.1$ and standard deviation 0.3}
\tablenotetext{g}{Blank spaces indicate no inference on the parameter as the posterior PDF was essentially uniform}
\end{deluxetable*}